\documentclass[fleqn,usenatbib]{mnras}

\usepackage{newtxtext,newtxmath}

\usepackage[T1]{fontenc}
\usepackage{ae,aecompl}
\usepackage{hyperref}
\usepackage{tabularx}

\newcommand{\kepler}{\textit{Kepler}}

\usepackage{graphicx}	
\usepackage{amsmath}	
\usepackage{comment}
\usepackage{tabularx}

\title[Generalized Exomoon Corridor]{The Exomoon Corridor for Multiple Moon Systems}

\author[A. Teachey]{Alex Teachey$^{1}$\thanks{E-mail: amteachey@asiaa.sinica.edu.tw}
\\
$^{1}$Academia Sinica Institute of Astronomy and Astrophysics \\ 
11F of AS/NTU Astronomy-Mathematics Building, No.1, Sec. 4, Roosevelt Rd, Taipei 10617, Taiwan, R.O.C.\\
}

\date{Accepted 2021 June 24. Received 2021 June 8; in original form 2021 April 7}

\pubyear{2020}

\begin{document}
\label{firstpage}
\pagerange{\pageref{firstpage}--\pageref{lastpage}}
\maketitle

\begin{abstract}
Recently \citealt{moon_corridor} identified the so-called ``exomoon corridor'', a potentially powerful new tool for identifying possible exomoon hosts, enabled by the observation that fully half of all planets hosting an exomoon will exhibit transit timing variation (TTV) periodicities of 2-4 epochs. One key outstanding problem in the search for exomoons, however, is the question of how well the methods we have developed under the single moon assumption extend to systems with multiple moons. In this work we use $N$-body simulations to examine the exomoon corridor effect in the more general case of $N \geq 1$ moons, generating realistic TTVs produced by satellite systems more akin to those seen in the outer Solar System. We find that indeed the relationship does hold for systems with up to 5 moons in both resonant and non-resonant chain configurations. Our results suggest an observational bias against finding systems with large numbers of massive moons; as the number of moons increases, total satellite mass ratios are generally required to be significantly lower in order to maintain stability, or architectures must be more finely tuned to survive. Moons produced in impact or capture scenarios may therefore dominate early detections. Finally, we examine the distribution of TTV periods measured for a large number of \kepler\ objects of interest (KOIs) and find the same characteristic exomoon corridor distribution in several cases. This could be dynamical evidence for an abundance of moons in the field, though we caution against strong inferences based on this result.

\end{abstract}




\section{Introduction}
As the search for exomoons continues, so too does the search for powerful new tools to identify them. For a variety of reasons, time-domain photometry remains our best bet for detecting exomoons at present: with missions like \kepler\ \citep{Kepler} and TESS \citep{TESS}, there is an abundance of data available for a broad moon search. Moreover, time-domain photometry can in ideal circumstances reveal three distinct but self-consistent signatures of exomoons: moon transits, transit timing variations (TTVs), and transit duration variations (TDVs) \citep[e.g.][]{Sartoretti:1999, Szabo:2006, Cabrera:2007, Kipping:2009, TDV-TIP}. 

In general, none of these signals on their own would be quite enough to claim an unambiguous moon detection; we would ordinarily want to see at least two of these signals providing corroborative evidence for a moon. For example, several putative moon transits with physically plausible associated TTVs could be convincing. Or, the combination of TTVs and TDVs, with the same period, expected phase shift \citep{Kipping:2009, TDV-TIP, Heller_MMR}, and amplitudes suggesting a common mass and semimajor axis solution, might be compelling, even if the moon's transit is in the noise or missing entirely. Unfortunately, TDVs are exceptionally difficult to measure due to their small amplitudes, and TTVs by themselves have many other possible causes. This is unfortunate, because TTVs can often be measured to high precision, so we would certainly like to be able to utilize this information, not only in the course of vetting moon system candidates, but also in identifying those candidates in the first place.

To that end, \citealt{moon_corridor} (hereafter K21) recently identified a phenomenon called the ``exomoon corridor'', in which exomoons, regardless of their underlying semimajor axis distribution, manifest predominantly short TTV periodicities. Because moon orbits will be undersampled -- the period of the moon $P_{\mathrm{S}}$ will always be shorter than the period of the planet $P_{\mathrm{P}}$ \citep{Kipping:2009} -- the periods we measure will always be aliases of the underlying oscillation signal.

This is potentially a very useful observation, not only because TTVs are relatively easy to measure, but also because this distribution, with 50\% of all exomoon systems showing periodicities of 2-4 epochs, appears distinct from the distribution of TTV induced by planet-planet interactions. For comparison, K21 computed TTVs for 90 known planet pairs (180 planets) investigated by \citealt{HL2017} (henceforth HL2017), and found that by contrast only 1\% of these systems display such short-period TTV solutions. Thus, if we happen to measure a TTV with a period between two and four cycles, this system is immediately of interest because planet-planet perturbation -- probably the least exotic explanation available -- is comparatively far less likely (though not 50 times less likely; see K21 for further discussion on this interpretation).

A lingering challenge, however, is the problem of systems with multiple moons. The single moon assumption has been built in to the K21 analysis, and indeed has been central to most exomoon photometric search efforts to date \citep[e.g.][]{HEKI, TK18, Alshehhi:2020, Fox:2021}. This is in part because the inclusion of each subsequent moon requires seven additional model parameters, and so as we add additional moons it very quickly becomes computationally demanding to compute Bayesian evidences. Moreover, we quickly enter a regime in which the data are simply insufficient to support such a complex model.

This is problematic, because while the system is formally, \textit{mathematically} more complex, it is not really more complex in the sense of requiring significant additional credulity. That is, the presence of multiple moons in a system does not so strain the imagination that we require overwhelming evidence for it; it may be equally or even more probable \textit{a priori}. But in the prevailing framework we will penalize each additional moon for the extra complexity -- the data simply cannot support it, unless we can factor in the relative probabilities of these different architectures, which are at present poorly constrained. Thus, we may be biasing ourselves against the detection of multi-moon systems.

Furthermore, while the single moon case can be reasonably approximated as a nested-two body problem, the presence of multiple moons means there will be more complicated dynamical interactions at play -- not just multiple gravitational forces acting on the planet, but also mutual interactions of the moons with one another -- which can only be simulated properly through $N-$body integrations. A simple model, comprised of a superposition of multiple sinusoids, could of course be computed analytically for any configuration of moons, but the stability of these systems is not a given. We must produce a realistic sample of these systems if we hope to make inferences about the population.

One notable exception to the single moon paradigm in the literature was \citealt{Heller_MMR}, who examined the effect multiple moons would have on both TTVs and TDVs, and found that these systems certainly do look significantly different than single-moons systems in TTV-TDV diagrams. While in the single-moon case, a moon will trace out an ellipse in TTV-TDV space, multiple moons will trace a more complex morphology. Unfortunately, the authors concluded these patterns would not be detectable with current instruments, largely due to the aforementioned low signal-to-noise ratio (SNR) for TDVs with current facilities. 

In the short term, then, if we wish to use dynamical signatures to search for possible exomoon hosts, we will have to rely on leveraging TTVs by themselves. Hence the great promise of the exomoon corridor.

In this work we investigate whether the exomoon corridor finding holds for the more general case of $N \geq 1$ moon systems. Considering the additional dynamical complexity of multiple moons, it is by no means obvious that the single-moon assumption holds in the very plausible context of a planet hosting more than one moon. Of course, the majority of massive moons in our Solar System (those that are potentially detectable by an extrasolar \textit{Kepler} or \textit{Hubble} analog) are found in multi-moon systems, so it is quite reasonable to anticipate the same in exoplanetary systems.

To answer this question, we have produced $N$-body simulations of planet-satellite systems containing 1, 2, 3, 4, and 5 moons, computing the resulting TTVs and examining the distribution of their periods. In section \ref{sec:theory} we briefly discuss the background theory. Section \ref{sec:sims} describes building and carrying out the simulations, constructing the simulated observations, and performing quality controls. In section \ref{sec:results} we report our results, including a diagnosis of the stability of these systems and their detectability. Section \ref{sec:analysis} discusses the impact of the results and examines the distribution of TTVs for our simulated moon systems against the transit timings measured with \kepler\ photometry by \citealt{holczer:2016}. We conclude in section \ref{sec:conclusions}.

\label{sec:introduction}

\section{Theory}
\label{sec:theory}
The exomoon corridor result is based on the aliasing of moon signals as manifested in the Lomb-Scargle periodogram \citep{lomb:1976, scargle:1982}. For the exomoon case, the phenomenon arises because the barycentric oscillation of the planet induced by the moon's gravitational tug is always higher frequency than the sampling cadence \citep{Kipping:2009} which, because we are measuring transit timings, is dictated by the orbital frequency of the planet. The result is an undersampled signal with an aliased period given in K21 by 

\begin{equation}
    P_{\mathrm{TTV}} = \frac{1}{\frac{1}{P_{S}} - \mathrm{round}\left[ \frac{P_P}{P_S} \right] \frac{1}{P_P} } .
\end{equation}

\begin{figure}
    \centering
    \includegraphics[width=8.8cm]{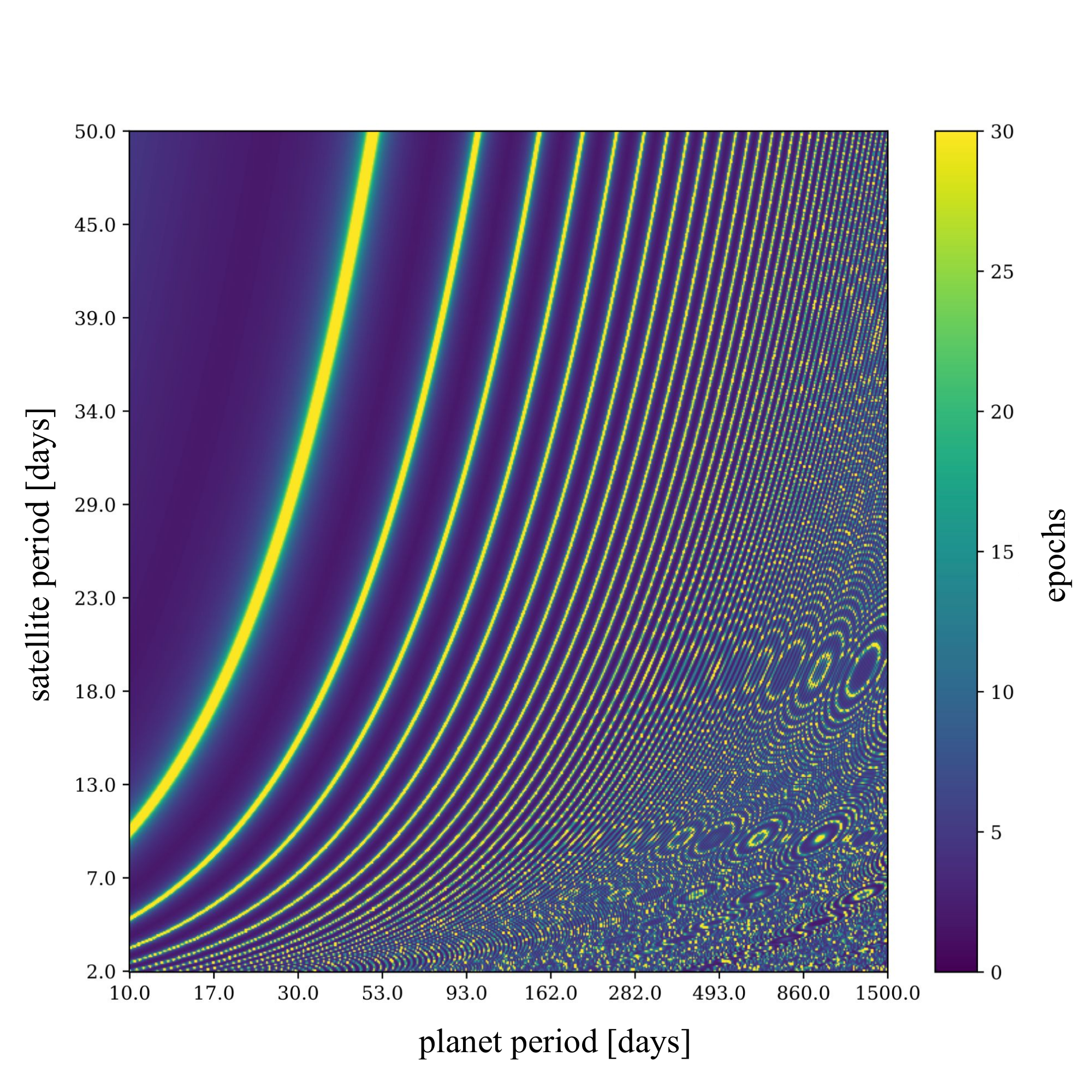}
    \includegraphics[width=8cm]{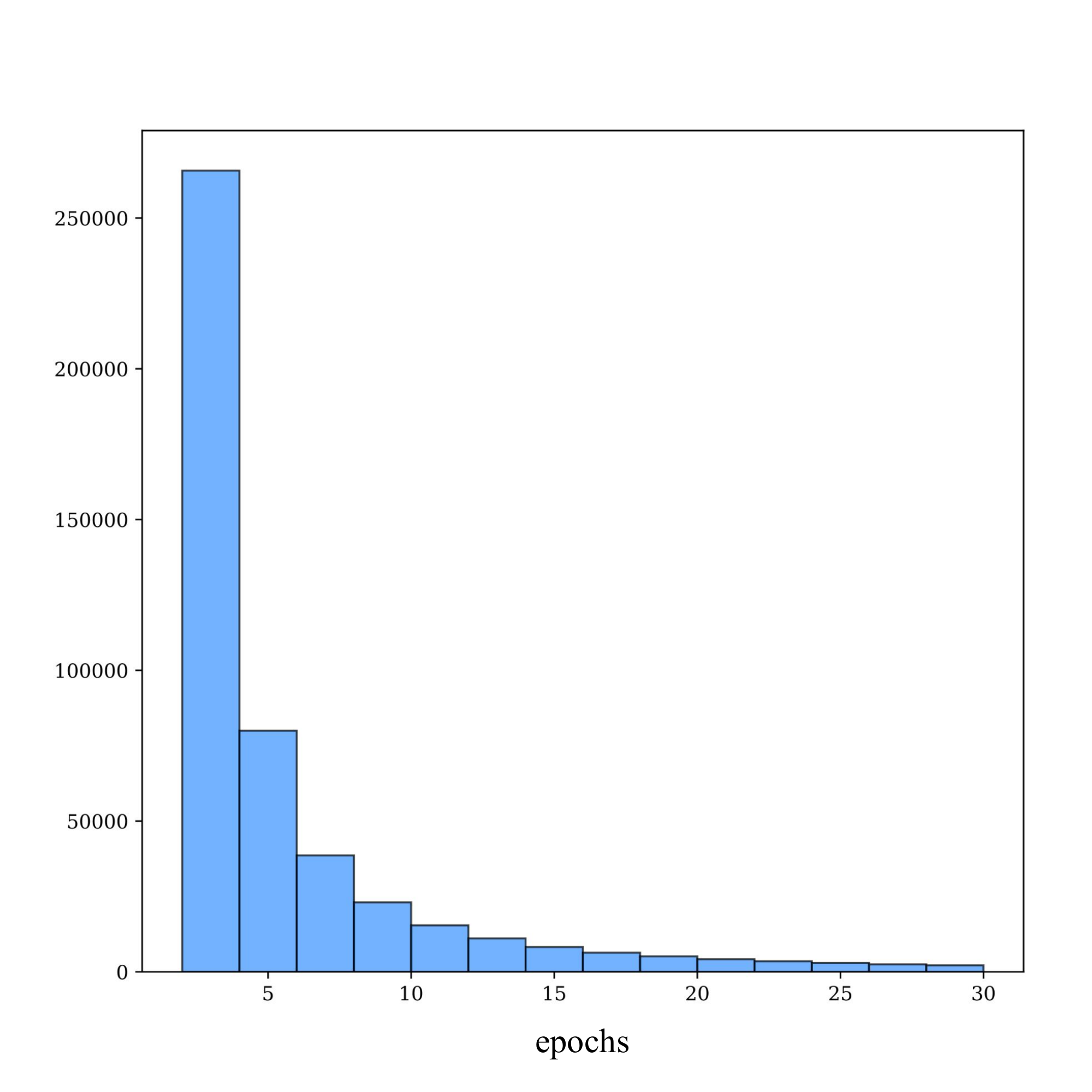}
    \caption{\textit{Top:} Aliased TTV periods as a function of moon and planet periods. The complex morphology towards the right of the plot appears to be a a moir{\'e} pattern resulting from the finite resolution of the grid (i.e. it is not real structure). \textit{Bottom:} a histogram of all the $P_{\mathrm{TTV}}$ values in the plot above. The pile-up at short periods is evident.}
    \label{fig:PTTV_aliases}
\end{figure}

\noindent With this straightforward expression we can produce Figure \ref{fig:PTTV_aliases}, which maps the aliased signal $P_{\mathrm{TTV}}$ as a function of the period of the planet $P_P$ and the satellite $P_S$. Below the $P_{\mathrm{TTV}}$ map we histogram the distribution of these aliased periods and find the characteristic pile-up at short periods found in K21. This simple picture then suggests that there is an exact solution to this problem, and that we might even be able to back out the true period of the moon from a given measurement of $P_{\mathrm{TTV}}$ and $P_P$. Unfortunately, as K21 points out, and as the map shows, there are effectively infinite $P_S$ solutions for a given value of $P_P$ that will produce the observed $P_{\mathrm{TTV}}$. 

It would be possible to break this $P_S$ degeneracy, at least to a point; there are not really infinite choices for $P_{S}$. We could do better by modeling the stability of the system in question to determine plausible $P_S$ solutions, establishing a possible mass range for the planet and/or moon based on transit depths (or upper limits on size based on non-detections), and / or incorporating the measurement of (or upper limits on) TDVs. This is demanding work, but there is no getting around these additional efforts to better constrain $P_S$, it simply does not fall out of the equation above.

Further complicating the matter is the fact that the analytical $P_{\mathrm{TTV}}$ formula above cannot account for the presence of \textit{multiple} aliases or harmonics within the periodogram. These do not show up in the K21 analysis, as that work does not grapple with the confounding issues of missing transits (making the sampling irregular), and the presence of timing uncertainties. For real observations these will certainly be a factor.

The system clearly has one ground truth solution for both $P_P$ and $P_S$, but Lomb-Scargle periodograms typically manifest several peaks associated with a single signal, arising from the finite sampling of the signal. For a well- and uniformly-sampled signal with little or no noise, these peaks are typically, but not always, strongest at the fundamental frequency. On the other hand, the peak power may not be at the fundamental frequency if the signal is not composed of a single sinusoid \citep{VP_LS}. This is precisely the case for a system with $N > 1$ moons, where multiple signals will be superposed.

In the case of noise-free signals and uniform sampling, the periodogram of $N$ signals combined will be roughly equivalent to the sum of the periodograms for each individual signal, divided by some factor $N - 1 \leq a \leq N$. This clean relationship begins to break down when photometric noise is introduced, and especially, when sampling becomes non-uniform. Fortunately, in the case of measuring TTVs, the sampling cadence will be quite regular, as it is dictated by the orbital period of the planet, not on stochastic observational opportunities (as in the case of performing ground-based observations of other astrophysical events in the time-domain). Even so, the loss of even a small number of observations, which may arise from (for example) the transit occurring during a temporary data gap, can introduce significant additional power into the periodograms. These are of course not astrophysical in nature, but they will complicate our analysis. Combine this with the presence of an unknown number of signals and the situation can be quite murky indeed.

In the case of real data, then, in the presence of multiple signals, it is not at all straightforward to predict analytically how the periodogram will manifest these signals, and indeed, they are famously challenging to interpret \citep[see][]{Dawson:2010, VP_LS}. In light of this, the safest and most comprehensive approach to answering this question of whether the exomoon corridor result generalizes to $N > 1$ moons is to simulate real systems and test it directly.


\section{N-body Simulations}
\label{sec:sims}


\subsection{Sample design}
To produce the sample, we built three sets of 50,000 systems (150,000 total) with the $N$-body simulator \texttt{REBOUND} \citep{REBOUND}, using the \texttt{WHFAST} integrator \citep{WHFAST}. Each system contained one, two, three, four, or five moons, the generation of which we describe momentarily. 

The first set of simulations consisted of planets of uniform (Jupiter) mass, all orbiting a Solar analog, with a consistent photometric uncertainty per observation of 350 ppm (roughly that of a 15th magnitude star as observed by \kepler). This run serves as a basis of comparison, for which the observational properties (i.e. transit depth and stellar magnitude) are not a factor, so that only the architecture of the satellite system, and the orbital period of the planet, potentially play a role in the final detectability of moons. We call this the ``fixed host'' sample.

In the second set of 50,000 planets, we introduced additional variables so as to more closely mimic the real \kepler\ sample. A star was drawn at random from the set of confirmed and candidate planet hosts listed on the NASA Exoplanet Archive, from which the mass and radius would be taken to compute the planet's transit duration and region of stability. To produce a realistic planet, we drew (separately from the star draw) a ratio of radii $R_P / R_{*}$, and with that computed the radius of the simulated planet, utilizing the selected stellar radius from before. The associated photometric uncertainty for this planet draw was also recorded so that it could be utilized to calculate the transit timing uncertainty. A realistic mass for this planet was then generated using the empirical mass-radius code \texttt{forecaster} \citep{forecaster}. We call this the ``variable host'' sample.

For the final set of simulations, we again utilized the variable host framework for system generation, but further required that every moon be initialized in resonance with its neighboring interior moon. The process of selecting these orbits is described momentarily.

For all three sets of simulations, planet orbital periods were chosen from log-uniform distribution between 10 and 1500 days. From these values we could compute the Roche limit $R_{\mathrm{Roche}}$ and Hill radius $R_{\mathrm{Hill}}$ for each planet. Note that we did not directly simulate the gravitational influence of the star on the planet-moon system within \texttt{REBOUND}; the stellar properties are used simply to compute the stability limit of the planet, and the photometric properties of the simulated observation -- namely, the transit durations and timing uncertainties, which we discuss below.

For each simulation, we selected a total moon system-to-planet mass ratio from a uniform distribution between $10^{-5}$ and $10^{-2}$ (roughly the range of total moon masses observed in the Solar System). The number of moons for this system was also selected at random (between 1 and 5), and with this number in hand the masses of each moon were chosen randomly and then modified iteratively until their combined mass added up to the desired total mass. For example, if four moons were to be generated for a given system, we started by drawing four mass percentages at random, each between 10 and 90 percent. Upon first generation these percentages are very unlikely to  add up to 100\%, so from here each percentage would be raised or lowered iteratively by 2\% of the said value until they did sum to 100\%. In this way, we produced a randomly generated sample of multi-moon systems, some of which may have roughly similar masses, others with quite disparate masses across the system, and with a variety of architectures (massive moons on the inside, or towards the outside, or interspersed with less massive moons). Because we draw these values randomly, the initial ordering of these moon masses may not necessarily reflect established pathways of moon formation, but we wanted to be agnostic about such pathways, as we may yet encounter systems in nature that have not been anticipated. Instead, we let the stability of the systems be our guide for determining which of these architectures may be plausibly found in nature. 

After the number of moons was chosen, the stability region for prograde moons, between $R_{\mathrm{Roche}}$ and $0.4895 \, R_{\mathrm{Hill}}$ \citep [][]{Domingos:2006} \footnote{More recently \citealt{Rosario-Franco:2020} found a narrower stability range, with a critical Hill sphere fraction $f_{
\mathrm{crit}} = 0.4061 \, R_{\mathrm{P}}$, and stability up to the 
\citealt{Domingos:2006} limit utilized here in only a fraction of cases. In this context, adopting a narrow stability range for prograde moons only affects the median amplitude of induced TTVs. Our results are unchanged by enforcing a narrower stability region.}, was divided into $N$ equal-sized segments (the number of segments equaling the number of moons), and the first moon $S_1$ ($i = 1$) was placed randomly somewhere within the innermost segment. For each subsequent moon (in the non-resonant simulations), the remaining stability region was re-divided into $N - i$ segments, with the new inner limit becoming 150\% the previous moon's semimajor axis, and the new moon's semimajor axis was once again selected at random to be somewhere within the new innermost segment. Thus moons were spaced evenly but not uniformly across the stability region. After some experimentation we found this approach minimized the number of systems that had to be rejected during the generation process, and ensured that the moons were numbered in order of their distance from the planet. Ultimately we elected to space half our systems drawing from a uniform distribution and the other from a log-uniform distribution to further diversify the architectures.

For the systems initialized in period resonance, the first moon semimajor axis was selected as before (i.e. placed randomly within the innermost segment). For each subsequent moon, a period ratio $b:a$ was selected randomly such that $a \in \{1,2,3,4,5\}$ and $b \in \{ a+1, a+2, ..., 4a\}$. Thus these period ratios are mostly randomized except there will be a pile-up at the reducible ratios (for example, 2:1 will have more representation due to the inclusion of 4:2, 6:3, 8:4, and 10:5; 3:2 will also include 6:4 draws, etc). Additional resonances may of course exist between two non-neighboring moons but these were not selected for or against.

Moon inclinations were all set to zero (co-planar orbits) to simplify calculations, in particular, to avoid transit duration variations induced by a changing impact parameter \citep{TDV-TIP}. Note that systems where a moon is inclined with respect to the orbital plane of the planet will produce the same TTVs as a system seen edge-on, so the only other effect we are leaving out due to this choice is the mutual interaction of moons that have orbits inclined with respect to one another, which would of course affect their stability. After some experimentation with non-zero moon eccentricities up to 0.01 (marginally larger than the eccentricity of Europa), we elected to set all initial eccentricities to zero, with the expectation that the interactions between the moons would excite these eccentricities naturally in some cases (and indeed they did). Starting phases for each moon were randomly selected between 0 and $2\pi$.


\subsection{System stability}
We did not wish to run lengthy (computationally expensive) simulations for each system; to produce a sample of 150,000 systems integrated over a typical $10^9$ orbits would require millions of CPU hours, and would be hard to justify for an experiment of limited scope. The central purpose of this work is not to explore the evolution of multi-moon systems exhaustively; rather, it is to observe the effect of plausible multi-moon systems on the measured TTVs to test the exomoon corridor prediction. Even so, the stability of these systems remains central to the question of whether the exomoon corridor holds in the more general case. We certainly do not want to include systems that cannot be found in nature due to dynamical instability, as this could severely bias our results, and therefore we turn to long-term stability proxies.

For the single moon case, long-term stability is expected, as we employ no other external perturbing forces. These forces are no doubt important in real systems \citep[e.g.][]{Quarles:2020}, but including stellar influence would make the simulations significantly more complex, and would ultimately prevent us from employing the machine learning tools we utilize in this work (see below).
Single moons may also experience orbital decay due to tidal forces or due to a non-spherical potential. We included the $J_2$ quadrupole moment in our simulations using the expansion package \texttt{REBOUNDx} \citep{REBOUNDx}, selecting from a uniform distribution between 0.001 and 0.016 (roughly the range found amongst the planets in our Solar System). However, in practice we found this had no effect on the simulations, as all moons were initialized with zero inclination and never left the $x-y$ plane. In any case, for the single moon case the TTVs are well modeled by a single sinusoid, so we are comparatively less concerned with how these systems manifest observationally, except as a point of comparison for $N>1$ systems.

For every system, regardless of the number of moons, we calculated the Mean Exponential Growth Factor of Nearby Orbits (MEGNO) number, a chaos metric, as a proxy for long term stability \citep{MEGNO}. For stable systems MEGNO tends towards the value of 2, while unstable systems generally take larger numbers. This number is calculated natively by \texttt{REBOUND}. For systems with two moons, we use the MEGNO number as the primary predictor of long-term orbital stability, and establish a threshold for stability as described below. 

\begin{figure}
    \centering
    \includegraphics[width=9cm]{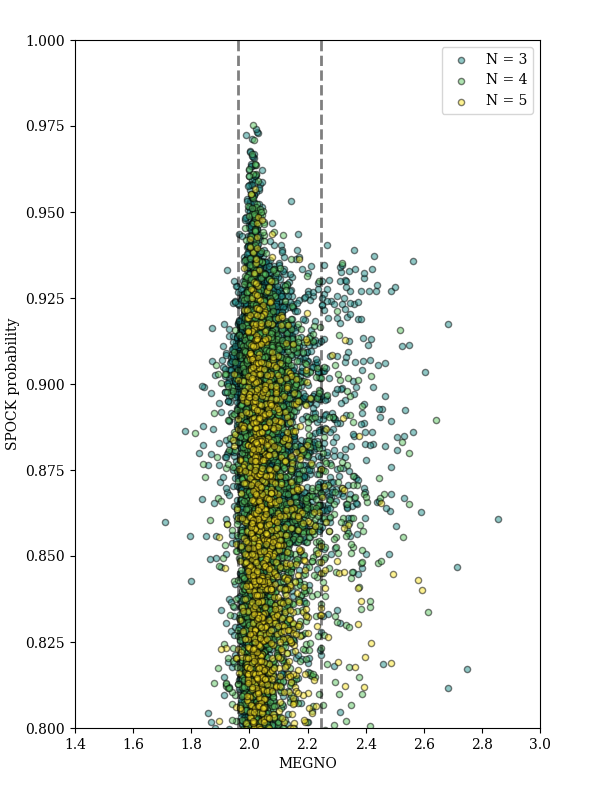}
    \caption{The \texttt{SPOCK} stability probability vs MEGNO value, for all systems in the fixed host simulations showing stability probability $\geq 80\%$. The distribution for the variable host simulations was similar. Note that only systems with 3 or more satellites are assigned a \texttt{SPOCK} stability probability. The range of MEGNO values considered stable is indicated by the dashed lines.}
    \label{fig:SPOCK_megno_limits}
\end{figure}

Finally, for systems with three or more moons, we employed the Stability of Planetary Orbital Configurations Klassifier (\texttt{SPOCK}) package, a powerful new tool which utilizes a machine learning framework to predict long term stability ($10^9$ orbits of the innermost satellite) from far shorter integrations \citep{SPOCK}. This package was produced to characterize the stability of planetary systems, and was trained on 3-planet systems, but has been shown to generalize well to $>3$ planet systems in both resonant and non-resonant orbits (it does not, however, provide stability predictions for systems with fewer than 3 moons). \texttt{SPOCK} provides an estimate for the probability of long-term stability, and we used these probabilities in tandem with the MEGNO number to select an appropriate cut-off for the latter in the systems with two moons (see Figure \ref{fig:SPOCK_megno_limits}). 

Each simulated system was normalized to be in agreement with the original \texttt{SPOCK} inputs, such that the mass of the host planet equals unity, as does the semimajor axis of the innermost moon. Every simulation was run for 10 Earth years, or roughly twice the time baseline of the \textit{Kepler} mission, to generate transit timings for these planets. This corresponds to $\mathcal{O} (10^3 - 10^4)$ orbits for the moons, depending on the semimajor axis (note however that \texttt{SPOCK} always carries out an integration of $10^4$ orbits for the innermost moon to produce its prediction). Only systems that ejected one or more satellites, or pulled a satellite within the Roche limit, were rejected during the generation process; all others were kept so that we could determine afterward the most appropriate metrics for exclusion. 

\begin{figure}
    \centering
    \includegraphics[width=8.8cm]{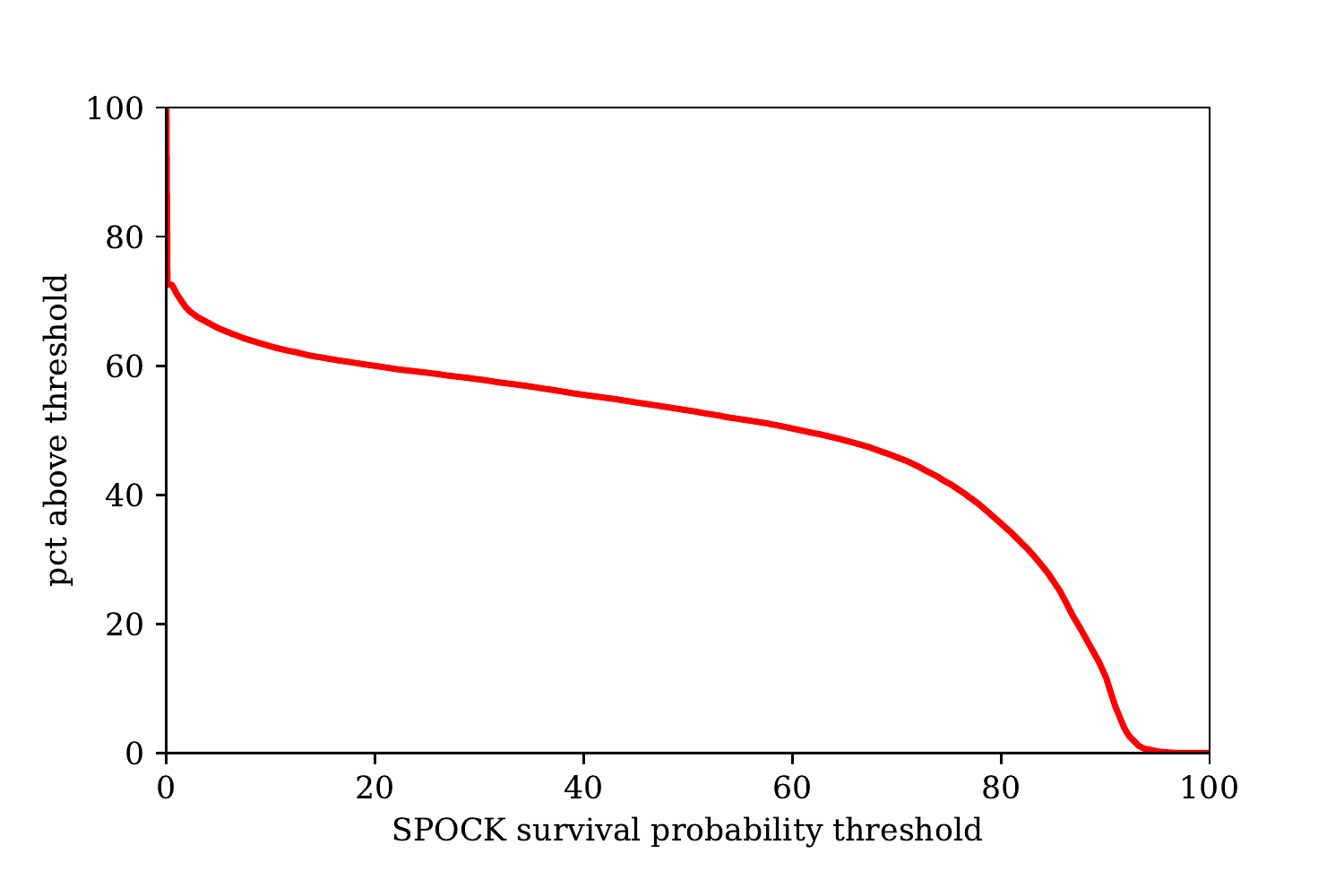}
    \caption{Percentage of systems with a \texttt{SPOCK} survival probability at or above a given threshold.}
    \label{fig:pct_above_spockprob}
\end{figure}

We found that roughly $15\%$ of our simulations with three or more moons resulted in a \texttt{SPOCK} simulation with survival probability of $\geq \, 90\%$ (see Figure \ref{fig:pct_above_spockprob}), so we adopted 90\% survival probability as our standard by which to determine a reasonable MEGNO value. Very few systems have $\geq 95\%$ stability probability, and none have $100\%$ probability (see Figure \ref{fig:SPOCK_megno_limits}), so we must make a trade-off between confidence in long-term stability and sample size. 

For these predicted stable systems, we compute a $2 \sigma$ range for MEGNO of [1.96, 2.25] for the fixed host simulations, [1.96, 2.22] for the variable host simulations, and [1.96, 2.09] for the resonant chain systems. These became our thresholds for stability in the two-moon systems. For systems where we have both a \texttt{SPOCK} probability and MEGNO value in hand, we only use the former, as it incorporates additional metrics (10 inputs in all) and because a chaotic MEGNO value does not necessarily mean a system will go unstable in practice \citep{SPOCK}. Nor does a good MEGNO number ensure stability, as Figure \ref{fig:SPOCK_megno_limits} shows that reasonable MEGNO values are also possible for systems where \texttt{SPOCK} predicts lower survival chances. In light of this, we urge caution in interpreting stability differences between $N = 2$ and $N \geq 3$ systems.

For each simulation we recorded the initial conditions, the $x-$, $y-$, and $z-$positions of each object in the system, their eccentricities and inclinations during the course of the simulation, the transit timings, and periodograms of the TTVs (description below). We also produced a summary file which contained for each system 1) the number of moons, 2) the planet's orbital period, 3) the number of transits observed, 4) the total satellites-to-planet mass ratio, 5) the TTV root mean squared (RMS) amplitude, 6) the MEGNO number, and 7) the \texttt{SPOCK} survival probability (where available).


\subsection{Computing transit times}
We computed transit times under the assumption that, in the absence of external gravitational effects (e.g. another perturbing planet), the planet-moon system barycenter travels on a Keplerian orbit with a period that does not change over the course of the observation time baseline. Thus, we can expect the barycenter should always transit like clockwork with the fixed period established for the simulation. We may then compute the physical displacement $\vec{r}_{\mathrm{disp}}$ of the planet from the barycenter at the time of transit, the $x-$component of which will be the the sky-projected lead or lag of the planet from the barycenter from the observer's point of view. Each moon starts with an orbit in the $x-y$ plane, which is co-planar with the planet's orbit; any $z-$ component to the orbit would mean an induced inclination, but in practice the simulations all remained co-planar.

The sky-projected velocity of the system in the direction of motion will then be given by $v_{x,\mathrm{obs}} = 2 R_{*} / T_{\mathrm{dur}}$, where $R_{*}$ is the stellar radius and $T_{\mathrm{dur}}$ is the duration of the transit (we set impact parameter $b=0$ for all simulations). Hence, the time delay from transit midtime of the barycenter $\tau_{\mathrm{bary}}$ and the transit midtime of the planet $\tau_{\mathrm{P}}$ will be given by $\Delta t = \tau_{\mathrm{bary}} - \tau_{\mathrm{P}} = x_{\mathrm{disp}} / v_{\mathrm{obs}}$. This value will be negative when the planet trails the barycenter from the perspective of the observer ($\tau_{\mathrm{bary}} < \tau_{\mathrm{P}}$). 

With an array of $\Delta t$ values in hand, we now have \textit{ground truth} TTVs, representing the absolute deviation of every transit time from ground truth linear ephemeris. However, because we have timing uncertainties and a finite baseline, the ground truth TTVs will not always be the same as what we would actually derive for a real system, because the true period of the barycenter transit is unknown to the observer. We therefore need to produce ``raw'' transit timings, by letting

\begin{equation}
    \tau_{m,\mathrm{obs}} = \tau_0 + m P_{\mathrm{gt}} + \Delta t_m 
\end{equation}

\noindent and then fit a line to this time series to \textit{infer} linear ephemeris. Here $\tau_m$ is the transit midtime for epoch $m$, $P_{\mathrm{gt}}$ is the ground truth period of the planet, and $\Delta t_m$ is the time displacement from barycenter at each transit midtime. We applied Gaussian noise to the raw transit times according to the relationship in \citealt{holczer:2016}, who computed an empirical relationship for transit timing uncertainty $\sigma_{\mathrm{TT}} = 100 \, \mathrm{min} / \mathrm{SNR}$. We verified that this empirical formula is in good agreement with the analytical formulae given in \citealt{carter:2008}, where $\sigma_{\mathrm{TT}} \approx (T_{\mathrm{dur}} / \mathrm{SNR})(0.5 R_P / R_{*})$ (with impact parameter $b$ and eccentricity $e$ both equal to zero, as implemented in our simulations). In general the empirical timing uncertainty is more conservative, so we adopted this for the simulations, and we further set an uncertainty floor of 6 seconds, which avoids extrapolating beyond the range presented by \citealt{holczer:2016}.

The resulting line fit is the inferred linear ephemeris, from which we can compute the TTVs (observed minus expected transit times); there is typically a small discrepancy between the ground truth orbital period and that inferred from the transit timings, but sometimes it may be appreciable.

\subsection{Computing the TTV period}
With the single moon case, in the absence of other perturbations and assuming a small eccentricity, it is reasonable to model TTVs as a single sinusoid. For multiple moon systems, the wobble of the planet will be more complex, as the moons will exert several gravitational influences on the planet, but we can model this also as simply the superposition of multiple sinusoids, and indeed this is functionally equivalent to modeling an arbitrary function with a Fourier series -- the approach we take to identifying periodicities. 

As the TTV amplitude scales linearly with both semimajor axis and moon mass \citep{Kipping:2009}, there is some degeneracy here in terms of which moons in a multi-moon system will dominate the signal. More massive moons at larger semimajor axes will of course dominate, as they will pull their common center of mass farthest from the planet's center; but the situation will be murkier if we have inner moons that are more massive than outer moons. For this work we make no assumptions about the system architectures other than building them from some reasonable distribution of masses, initializing orbits and requiring stability, and therefore there is the potential for a variety of system architectures in our final sample.

We proceeded then as we would under the single moon assumption, as that has been the standard to this point and also reflects the methodology of K21: with the simulation transit timings in hand, we utilize the Lomb-Scargle \citep{lomb:1976, scargle:1982} periodogram code distributed by \texttt{astropy} \citep{astropy1, astropy2}, specifying a period between 2 and 500 epochs, and testing 5000 logarithmically spaced periods in this range. These periodograms can have a range of morphologies, based on the presence of one or more sinusoids and also on the number and distribution of data points used \citep[c.f.][]{VP_LS}.

The peak power period for each periodogram was extracted and reported as the derived TTV period for the system. We then used this period as a fixed parameter for a least-squares sinusoid fitting to the TTVs, with only the amplitude and phase left as free parameters. We then computed the Bayesian Information Criterion \citep[BIC,][]{Schwarz:1978} as

\begin{equation}
    \mathrm{BIC} = k \ln{(n)} + \chi^2
\end{equation}

\noindent for both a flat line and the sinusoidal fit, where $k$ is the number of free model parameters, $n$ is the number of data points, and $\chi^2$ is the standard goodness-of-fit metric. We also computed the Akaike Information Criterion \citep{Akaike:1974} as

\begin{equation}
    \mathrm{AIC} = 2k + \chi^2
\end{equation}

Generally, the AIC is considered to be a better choice than the BIC for model selection in which the ``true'' model may not be among the tested options. Such is the case here, where we always fit a single sinusoid to the data but may very well be working with TTVs encoded with multiple sinusoids. We examined both metrics to see whether this choice might make a difference in the outcome. Ultimately we opt to use the $\Delta$BIC as we find it results in a more conservative selection of systems compared to the $\Delta$AIC.

While a single moon model is fully described by 14 parameters (compared to 7 for a planet-only model), for this model selection we are not actually fitting 7 additional parameters; we therefore computed the $\Delta \mathrm{BIC}$ and $\Delta \mathrm{AIC}$ here using an increase of only 2 free parameters (amplitude and phase), so $k=0$ for the flat line (no free parameters) and $k=2$ for the sinusoid. It is worth remembering that TTVs may have other causes besides the presence of a moon, so the safest approach is to ask simply whether a single sinusoidal TTV is favored over a flat line, not to require a physical explanation at this stage.

\begin{figure}
    \centering
    \includegraphics[width=7.4cm]{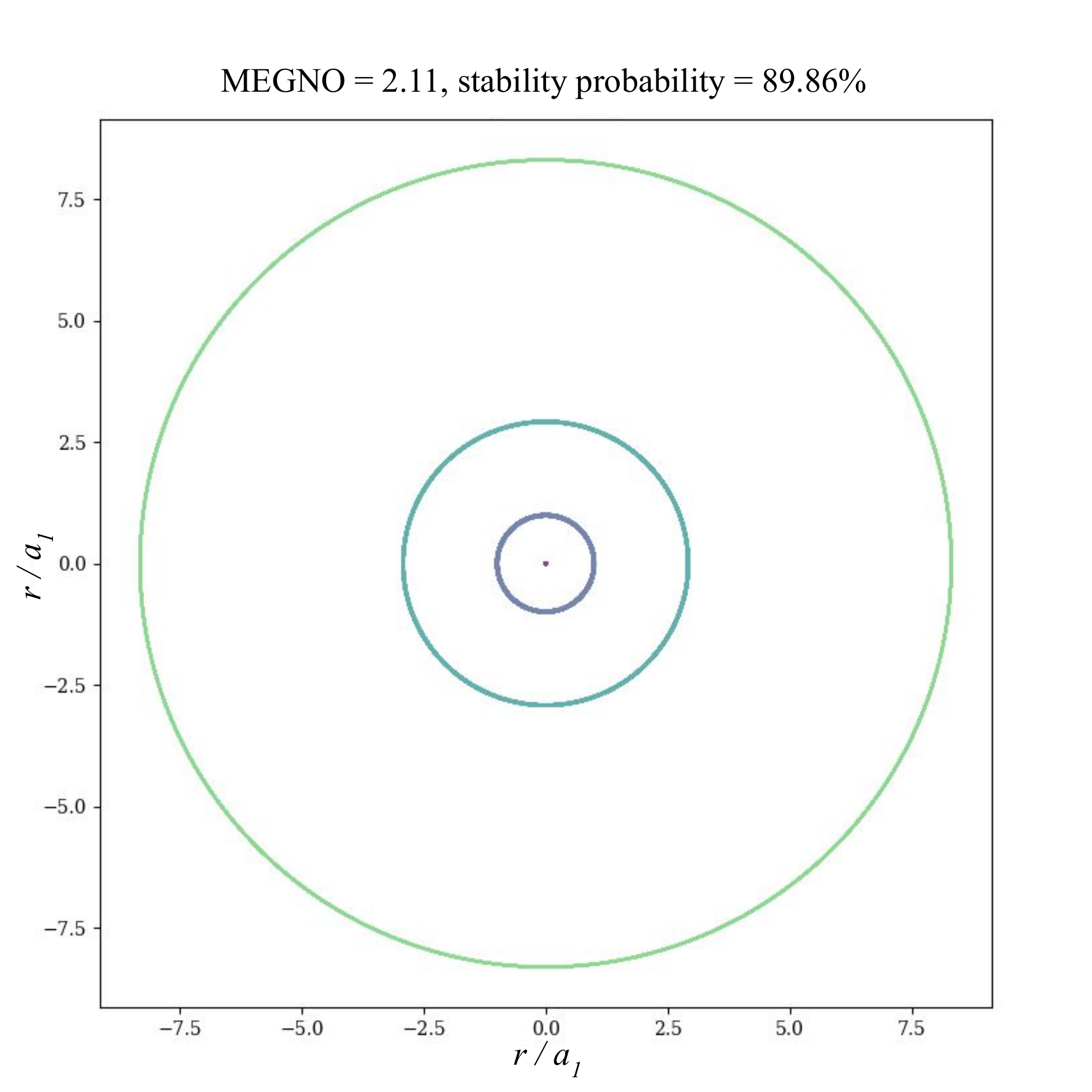}
    \vspace{1mm}
    \includegraphics[width=7.6cm]{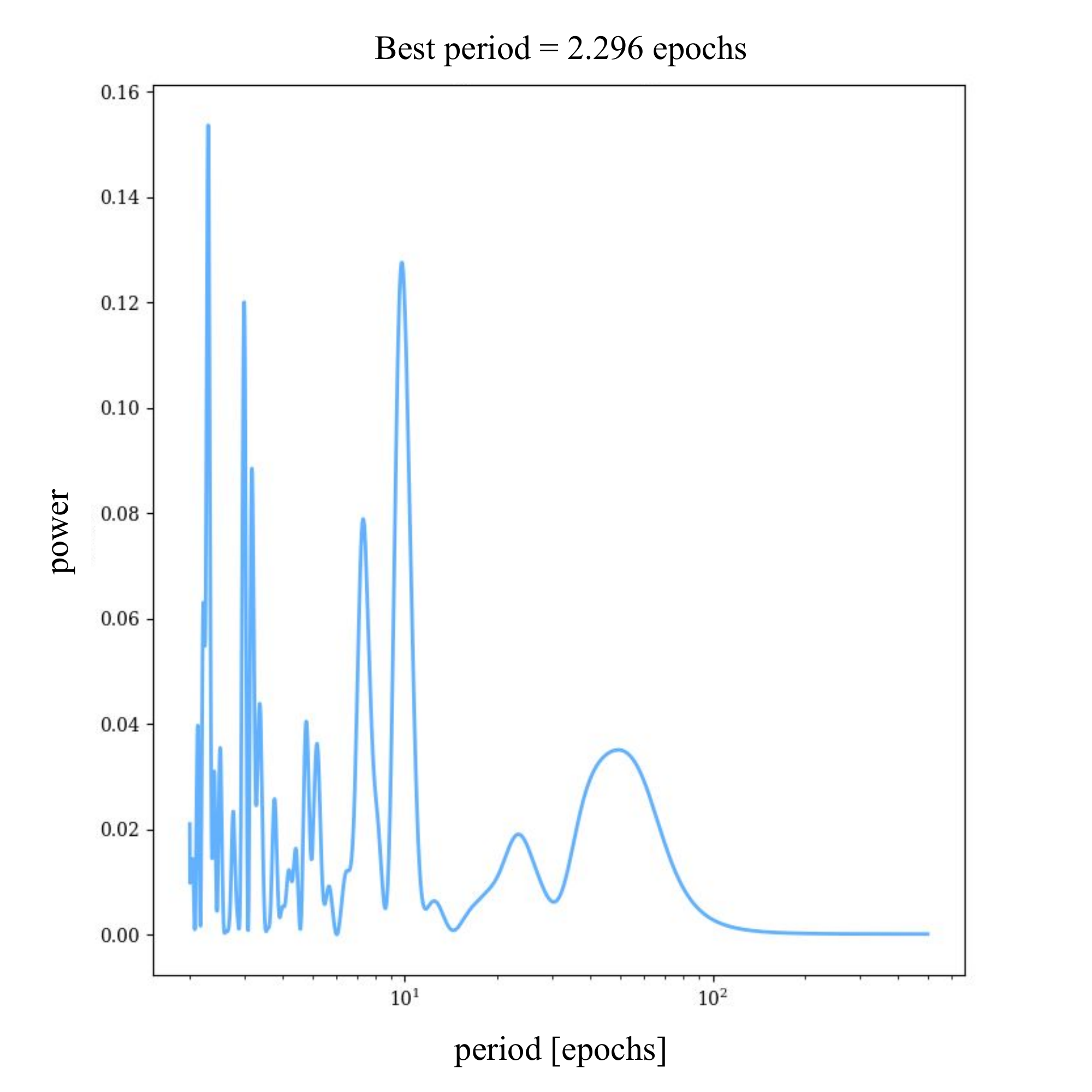}
    \includegraphics[width=7.6cm]{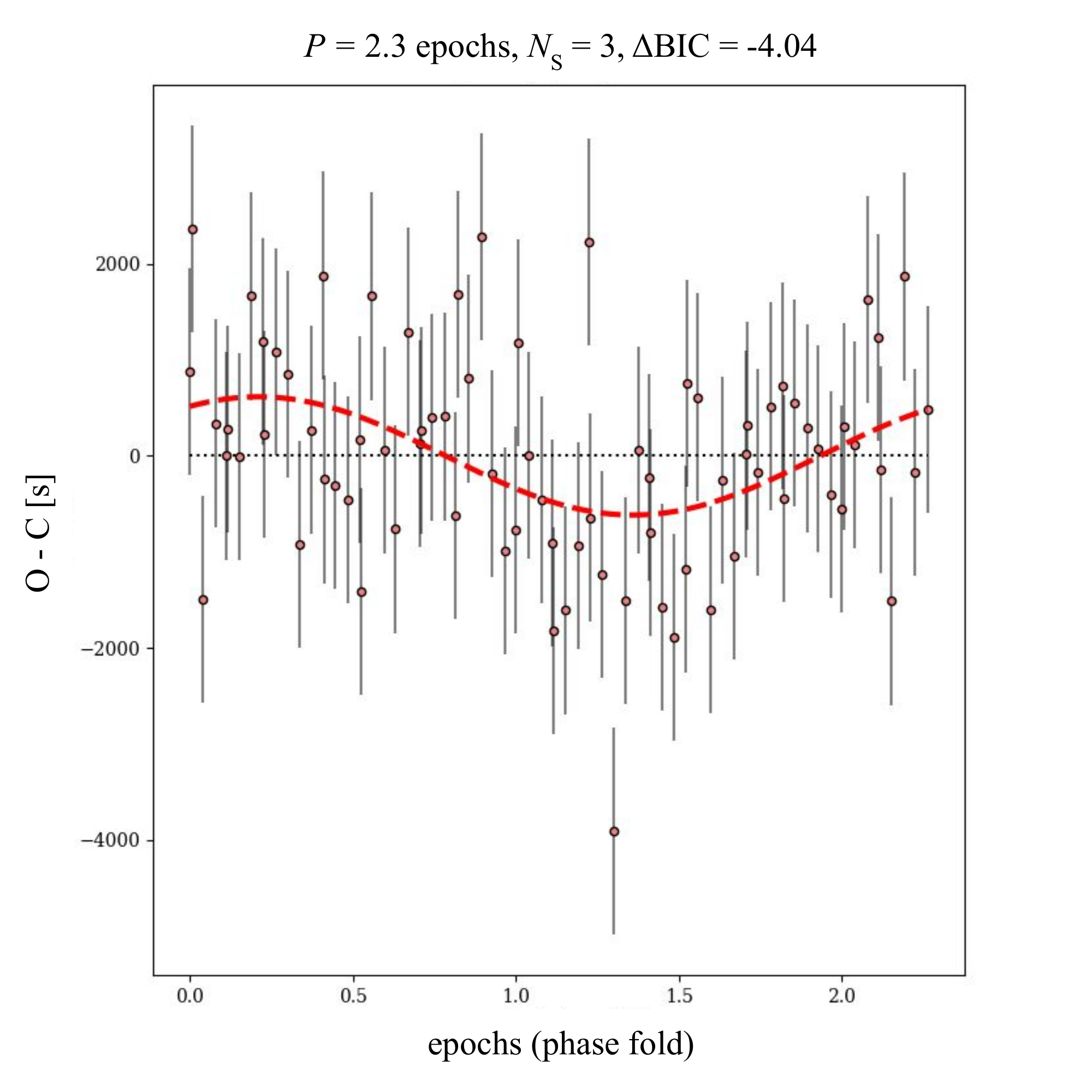}
    \caption{An example simulation. \textit{Top:} schematic of the system. \textit{Middle:} The periodogram run on the transit timings. \textit{Bottom:} The best fitting sinusoid to the transit timings.}
    \label{fig:example_simulation}
\end{figure}

\subsection{Cross-validating TTV solutions}
A simple $\Delta$BIC or $\Delta$AIC calculation can be valuable for identifying significant TTVs, but it is not an exceptionally robust metric for determining whether the competing model is correct; it merely characterizes a \textit{better} fit. In the present case, for example, we can imagine a range of sinusoidal fits (corresponding to peaks in the periodogram) that would yield evidence for a TTV based on these metrics, but in such a case we cannot be sure which of these periods is correct. Strictly speaking, there is no ``correct'' solution, both because we are measuring aliases, and because the power spectrum we produce has its shape by virtue of there being power at these frequencies. In other words, all these frequencies \textit{are} present, and therefore they are also correct.

Additionally, we note that individual outlier data points in a real sample -- especially those with insufficiently small uncertainties -- can have the effect of skewing a fit in such a way that high-frequency solutions are generally preferred, as higher frequency oscillations (possibly with larger amplitudes) will be more capable of fitting these spurious data points. This is a particularly worrisome phenomenon given the build-up of short period TTVs we are investigating in the present case. Therefore, because we are interested in the period solutions themselves -- not just testing the the presence of TTVs -- simply having a favorable BIC or AIC indicating a good fit is not enough. We want to measure a signal that holds up even if some of the data were stripped away. This test is also important as we will wish to compare the distribution of simulated TTVs to the real \kepler\ sample, and these also must be vetted for reliability. 

To this end, we carried out two additional steps for ensuring that our measured TTVs are robust. First, prior to the TTV fitting, we performed an outlier rejection on the transit timings using the \texttt{scikit-learn} implementation of the \texttt{DBSCAN} clustering algorithm \citep{DBSCAN} in one dimension (using the timing variations themselves; the spaces between epochs are not incorporated into the distance calculations). After some experimentation, we set the maximum neighborhood distance to be 5 times the median timing error, and the minimum number of samples per cluster at one fifth the total number of samples. These settings effectively screened out outliers in the \citealt{holczer:2016} catalog, which we use later to compare simulated TTVs with the real \kepler\ sample.

We then performed a cross-validation test consisting of up to 20 trials for each planet. For each trial, 5 percent of the transit timings were removed at random, after which the periodogram was re-run and a sinusoid was fit again using the peak period from this run. For any planet with fewer than 20 epochs present, the number of trials equalled the number of available epochs and only one data point was removed for each trial. The expectation behind this test was, for any real TTV signal present in the data, it should be robust against the removal of a small number of transit timings. By contrast, spurious periodic detections should be more incoherent, and thus the removal of transit timings will cause a shift in the inferred period. This test should also guard against choosing a periodogram solution that is produced by the observing cadence, since this will change for each run.

For each trial we recorded the median and standard deviation of the best fitting TTV period, the TTV curve amplitude, and the phase. Systems for which all of three of these metrics had a percent error of less than 5\% were considered reliable detections and used in our final analysis of the simulations and \kepler\ systems.


\section{Results}
\label{sec:results}

\begin{table*}
    \renewcommand{\arraystretch}{1.3}
  \centering
  \begin{tabular}{|c|c|c|c|c|c|}
  \hline
    \textbf{Number of moons} & \textbf{1} & \textbf{2} & \textbf{3} & \textbf{4} & \textbf{5} \\
    \hline 
    \hline
    Total (fixed host) & 14819 & 5157 & 11612 & 10028 & 8384 \\
    \hline
    Total (\%) stable & 14793 (99.8\%) & 4752 (92.1\%) & 2694 (23.2\%) & 739 (7.4\%) & 161 (1.9\%) \\
    + $\Delta\mathrm{BIC} \leq -2$ & 12300 (83.0\%) & 3834 (74.3\%) & 1986 (17.1\%) & 448 (4.5\%) & 76 (0.9\%) \\
    + cross-validated & 11835 (79.9\%) & 3668 (71.1\%) & 1862 (16.0\%) & 414 (4.1\%) & 70 (0.8\%) \\
    \hline
    + $\Delta\mathrm{AIC} \leq 0$ & 14033 (94.7\%) & 4527 (87.8\%) & 2460 (21.2\%) & 621 (6.2\%) & 121 (1.4\%) \\
    + cross-validated & 13433 (90.6\%) & 4311 (83.6\%) & 2311 (19.9\%) & 574 (5.7\%) & 112 (1.3\%) \\
    \hline
    $P_{\mathrm{TTV}}$ [epochs] & $3.41^{+4.73}_{-1.19}$ & $3.38\,^{+3.94}_{-1.15}$ & $3.25\,^{+3.52}_{-1.11}$ & $3.49\,^{+2.56}_{-1.48}$ & $3.04\,^{+3.71}_{-0.99}$ \\[3pt]
    \hline
    $A_{\mathrm{TTV}}$ [min] & $1.84^{+11.38}_{-1.37}$ & $1.37\,^{+6.85}_{-0.97}$ & $1.46\,^{+10.22}_{-1.0}$ & $1.47\,^{+15.27}_{-0.99}$ & $0.70\,^{+5.34}_{-0.36}$ \\[3pt]
    \hline
    $(\sum M_S) \, / \, M_P$ & $5.5^{+3.1}_{-3.3} \times 10^{-3}$ & $5.5\,^{+3.2}_{3.3} \times 10^{-3}$ & $3.9\,^{+3.7}_{-2.6} \times 10^{-3}$ & $2.7\,^{+3.6}_{-2.1} \times 10^{-3}$ & $6.7\,^{+31.9}_{-5.5} \times 10^{-4}$ \\[3pt]
    \hline
    \hline
    Total (variable host) & 15421 & 5148 & 11596 & 9486 & 7773 \\
    \hline
    Total (\%) stable & 15392 (99.8\%) & 4732 (91.9\%) & 2549 (22.0\%) & 550 (5.8\%) & 149 (1.9\%) \\
    + $\Delta\mathrm{BIC} \leq -2$ & 6418 (41.6\%) & 1905 (37.0\%) & 880 (7.6\%) & 173 (1.8\%) & 34 (0.4\%) \\
    + cross-validated & 6217 (39.9\%) & 1856 (35.6\%) & 827 (7.0\%) & 158 (1.6\%) & 34 (0.4\%) \\
    \hline
    + $\Delta\mathrm{AIC} \leq 0$ & 12489 (80.1\%) & 3812 (73.1\%) & 1788 (15.2\%) & 333 (3.5\%) & 87 (1.1\%) \\
    + cross-validated & 11803 (75.7\%) & 3619 (69.4\%) & 1666 (14.2\%) & 299 (3.1\%) & 83 (1.1\%) \\
    \hline    
    $P_{\mathrm{TTV}}$ [epochs] & $3.44\,^{+4.21}_{-1.15}$ & $3.42\,^{+4.06}_{-1.17}$ & $3.24\,^{+3.13}_{-0.98}$ & $3.3\,^{+2.16}_{-0.89}$ & $3.73\,^{+3.61}_{-1.36}$ \\[3pt]
    \hline
    $A_{\mathrm{TTV}}$ [min] & $2.96\,^{+5.87}_{-2.02}$ & $2.65\,^{+5.14}_{-1.79}$ & $2.8\,^{+5.06}_{-1.87}$ & $2.6\,^{+5.34}_{-1.71}$ & $2.37\,^{+7.69}_{-1.86}$ \\[3pt]
    \hline
    $(\sum M_S) \, / \, M_P$ & $5.6\,^{+3.1}_{-3.5} \times 10^{-3}$ & $5.5\,^{+3.2}_{-3.5} \times 10^{-3}$ & $3.3\,^{+4.2}_{-2.6} \times 10^{-3}$ & $1.5\,^{+4.6}_{-1.2} \times 10^{-3}$ & $4.4\,^{+9.7}_{-3.7} \times 10^{-4}$ \\[3pt]
    \hline
    \hline
    Total (resonant chain) & 15012 & 4839 & 12522 & 10251 & 7376 \\
    \hline
    Total (\%) stable & 14983 (99.8\%) & 4381 (90.5\%) & 1429 (11.4\%) & 455 (4.4\%) & 168 (2.3\%) \\
    + $\Delta\mathrm{BIC} \leq -2$ & 6125 (40.8\%) & 1529 (31.6\%) & 460 (3.7\%) & 127 (1.2\%) & 41 (0.6\%)  \\
    + cross-validated & 5833 (38.9\%) & 1471 (30.4\%) & 446 (3.6\%) & 125 (1.2\%) & 39 (0.5\%) \\
    \hline
    + $\Delta\mathrm{AIC} \leq 0$ & 12130 (80.8\%) & 3434 (71.0\%) & 1059 (8.5\%) & 301 (2.9\%) & 100 (1.4\%) \\
    + cross-validated & 11475 (76.4\%) & 3256 (67.3\%) & 1006 (8.0\%) & 284 (2.8\%) & 94 (1.3\%) \\
    \hline    
    $P_{\mathrm{TTV}}$ [epochs] & $3.42\,^{+3.98}_{-1.15}$ & $3.45\,^{+4.65}_{-1.17}$ & $3.49\,^{+3.61}_{-1.21}$ & $3.43\,^{+3.71}_{-1.23}$ & $4.05\,^{+2.26}_{-1.41}$ \\[3pt]
    \hline
    $A_{\mathrm{TTV}}$ [min] & $2.91\,^{+6.1}_{-1.99}$ & $2.38\,^{+4.78}_{-1.63}$ & $2.11\,^{+4.1}_{-1.45}$ & $2.32\,^{+4.61}_{-1.62}$ & $2.01\,^{+5.17}_{-1.51}$ \\[3pt]
    \hline
    $(\sum M_S) \, / \, M_P$ & $5.5\,^{+3.2}_{-3.4} \times 10^{-3}$ & $5.5\,^{+3.0}_{-3.7} \times 10^{-3}$ & $2.6\,^{+2.7}_{-2.1} \times 10^{-3}$ & $1.5\,^{+3.4}_{-1.3} \times 10^{-3}$ & $7.0\,^{+27.9}_{-5.2} \times 10^{-4}$ \\[3pt]
    \hline
    
 \end{tabular}
  \caption{Summary of the ``fixed host'' (top five rows) and ``variable host'' (middle five rows) and ``resonant chain'' (bottom five rows) simulations produced for this work. Percentages quoted indicate the fraction of total generated systems. Values quoted for $P_{\mathrm{TTV}}$, $A_{\mathrm{TTV}}$, and total satellite mass ratios are the median and 1$\sigma$ values for the robust cross-validation sample using $\Delta$BIC as the selection metric.}
  \label{tab:summary}
\end{table*}

\subsection{Stability}
Table \ref{tab:summary} offers a quick look at the results of our fixed host, variable host, and resonant chain simulations. The results from these three runs are quite similar in many respects, but we will highlight the differences where appropriate.

We start by noting that a large fraction of attempted two-moon systems were rejected during the generation process. For five architectures 
\textbf{($1 \leq N \leq 5$)} chosen randomly and 50,000 trials, we should have had roughly 10,000 samples each, but we ended up with just a little over half that number for the two-moon systems. Recall that ejection or fall-in were the only criteria for rejection during the simulation stage, so this particular outcome evidently occurs far more often than it does for three, four, and five moon systems with randomized orbits and masses. These other architectures, of course, also suffered ejections and fall-ins in some cases, but not to the same degree. We did not track how many times this occurred for each architecture, but at least 30\% of all attempted systems were rejected before completion.

\begin{figure*}
    \centering
    \includegraphics[width=18cm]{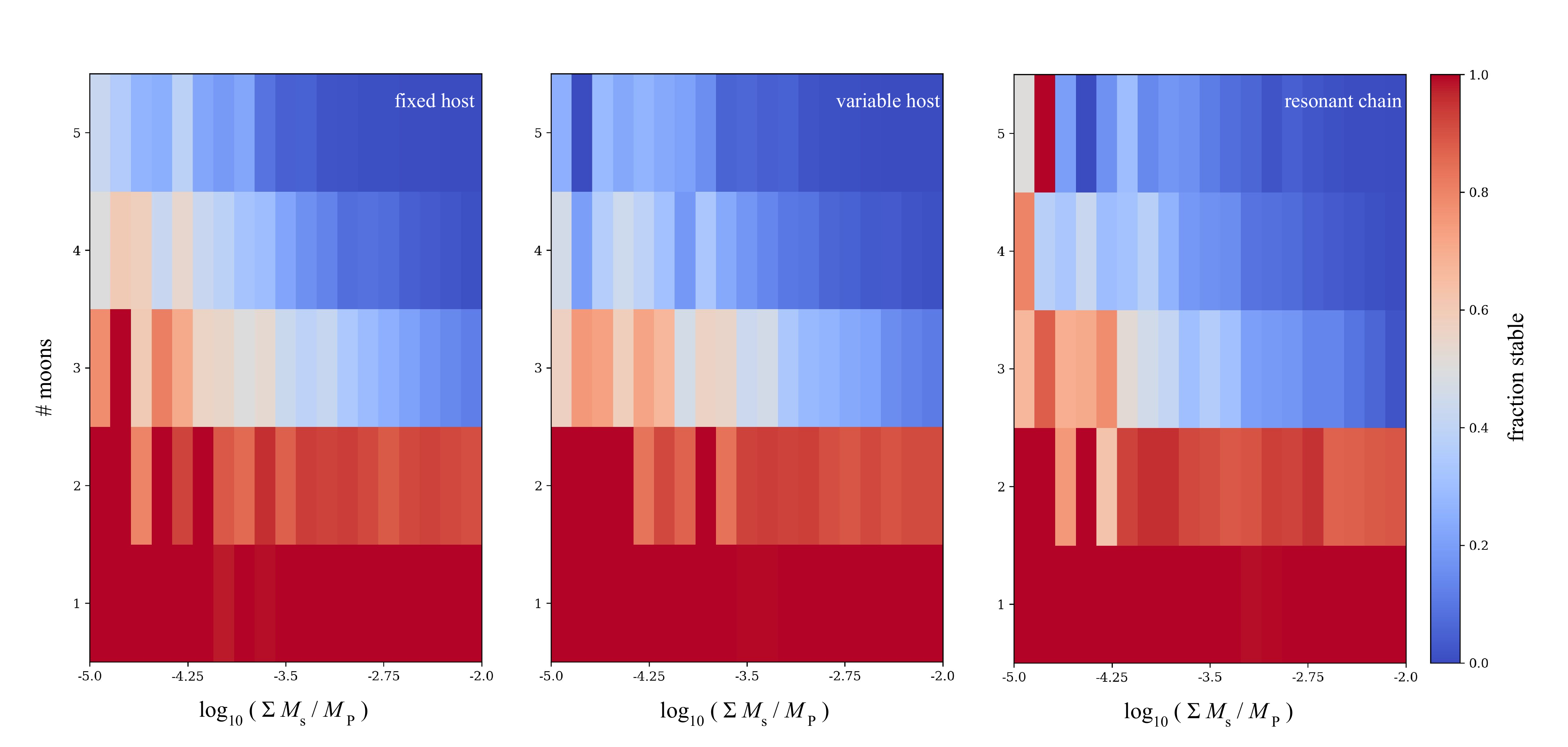}
    \caption{The fraction of stable systems for each architecture as a function of total satellite system mass in the fixed host (left), variable host (center), and resonant chain runs (right).}
    \label{fig:fraction_stable}
\end{figure*}

Thus, the 'total stable' numbers in Table \ref{tab:summary} and Figure \ref{fig:fraction_stable} should be interpreted as being the percentage of systems that are predicted to be stable after surviving the integration. In this context we can see that the vast majority of two-moon systems ($\sim 92\%$) in the simulated mass ratio range will survive so long as they have an architecture that is not effectively instantly unstable. This remains a useful metric, because it is safe to assume nature will not be making moons that are unstable on extremely short timescales. In other words, their configurations will of course not be random. Systems that can survive $\mathcal{O}(10^4)$ orbits on the other hand are at least plausibly formed by nature, even if they are eventually lost.

Figure \ref{fig:fraction_stable} shows the survivability of the moon systems simulated to completion as a function of total satellite system mass and architecture. For $N=1$ moon systems, they are virtually always stable; with no external perturbation, only strong tidal forces at small moon semimajor axes may disrupt these systems on the timescale of the simulation. Nevertheless about 0.2\% of these systems are eventually categorized as long-term unstable owing to a MEGNO number outside the established boundaries. Of course, more realistic simulations of single-moon systems, including the effects of stellar perturbation and moon inclinations, would reveal a non-spherical region of stability, but for our purposes (initializing all moon orbits as co-planar) this serves as a useful baseline of comparison.

Once we get to $N > 1$ moons, a trend becomes apparent: the fraction of initialized systems predicted to be stable falls off steeply as we increase $N$, and notably, the total mass of these systems is on average required to be significantly lower in order to remain stable. AT $N=5$, less than 2\% of the generated systems were predicted to be long-term stable. 

We point out that $N=2$ systems show the same median mass ratio as $N=1$ systems (see Table \ref{tab:summary}), which at first glance might be in tension with the trend just observed. This remains consistent however in view of the fact that we have opted to curtail the mass ratios of the satellite systems at $10^{-2}$. Considered in isolation, $N=1$ satellite systems could be stable up to a mass ratio approaching unity (a binary system). By contrast, we do not expect $N=2$ systems generally to be stable in that mass ratio regime. So there must be some mass ratio at which $N=2$ systems are majority unstable, consistent with the trend visible in Figure \ref{fig:fraction_stable}. Evidently that limit is above the mass ratio range we have explored here. Because for $N=1$ and $N=2$ systems we have only probed the parameter space where stability is basically uniform across the entire range, it is not surprising that the median value for $\sum M_{\mathrm{S}} \, / \, M_{\mathrm{P}}$, around $5 \times 10^{-3}$, is half that of the maximum value.

\begin{figure*}
\centering
\includegraphics[width=17cm]{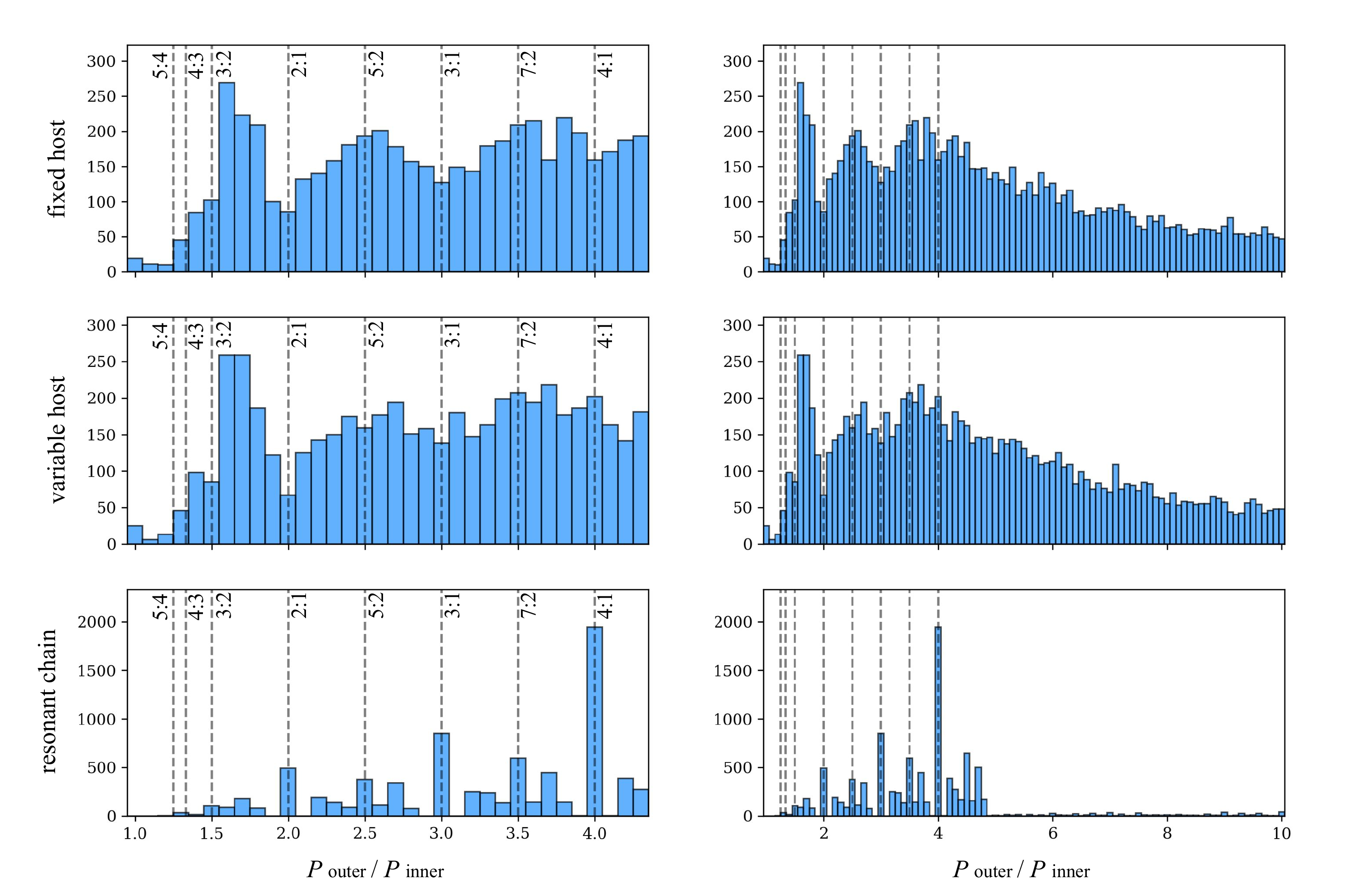}
\caption{\textit{Left:} Distribution of period ratios in the fixed host, variable host, and resonant chain simulations (detail). All moon pairs are included, not just adjacent pairs. \textit{Right:} the same as before but zoomed out to reveal more structure, though still with a truncated $x$-axis.}
\label{fig:period_ratios}
\end{figure*}

Figure \ref{fig:period_ratios} shows the distribution of period ratios for all three runs between all moon pairs. Slight deficits are evident around the resonances in both the fixed host and variable host runs, but of course the resonant chain run is comprised chiefly of moon pairs at these period ratios. 

We note also a reduction in the percentage of stable $N=3$ systems, by a factor of about 2, from the non-resonant to the resonant architectures. The cause of this is not obvious, and in utilizing a machine learning regressor with 10 inputs, a single culprit (for example, eccentricity growth) is not readily identifiable. No such reduction is apparent for the $N=4$ and $N=5$ systems, but this could be because resonance is no longer the filtering factor. As more moons are added, the moons are inherently more tightly packed, and finer tuning is required for stability. Hence, these stringent requirements for stability may be equally met in both resonant and non-resonant architectures.

\subsection{Detectability}

\begin{figure*}
    \centering
   \includegraphics[width=18cm]{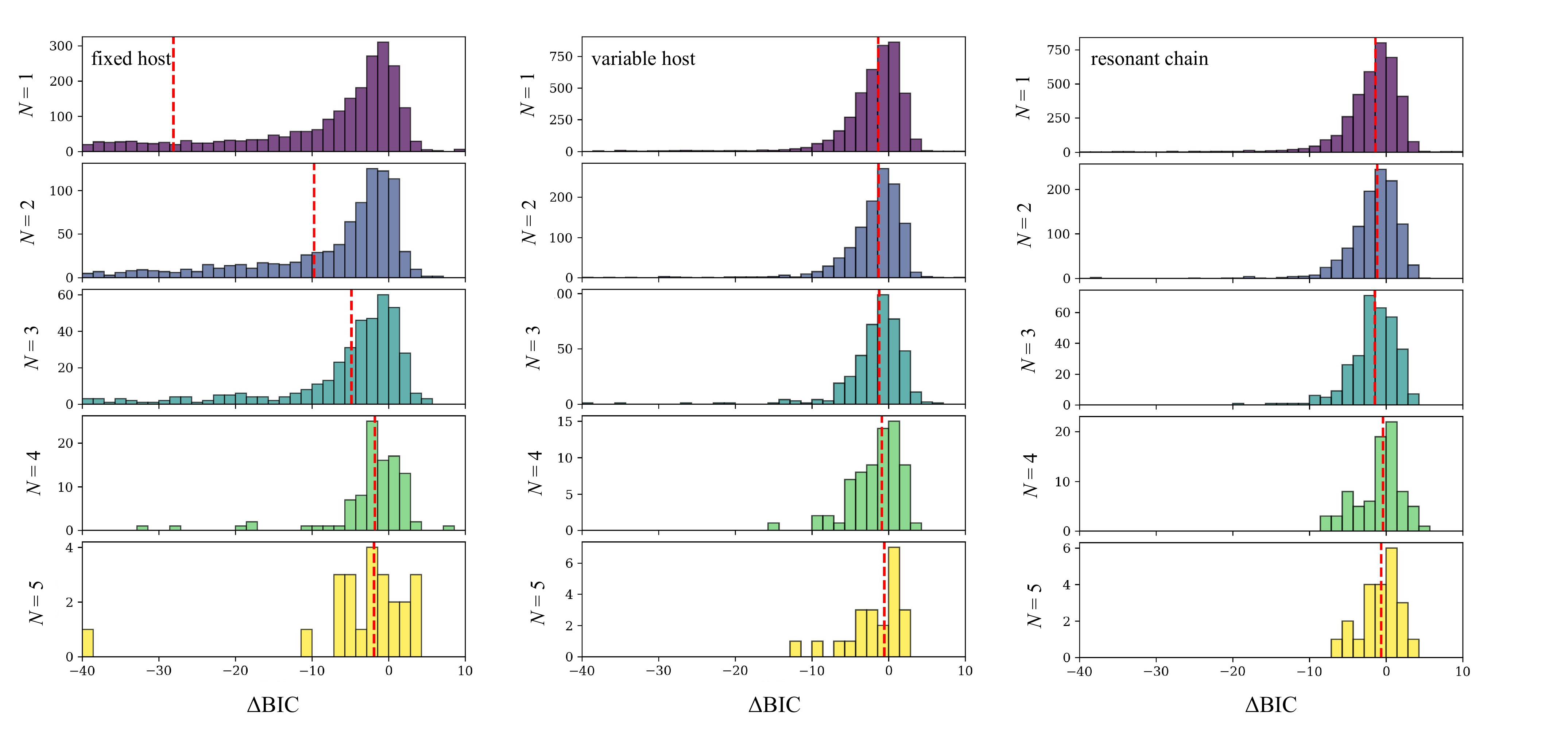}
    \caption{Distribution of $\Delta$BIC values for each architecture, for fixed host (left), variable host (center), and resonant chain (right) simulations. Representation in this sample is weighted by geometric transit probability. The red dashed line in each plot shows the median value of the distribution. The long distribution tails in the fixed host case are not shown.}
    \label{fig:deltaBIC_histograms}
\end{figure*}

\begin{figure*}
    \centering
    \includegraphics[width=18cm]{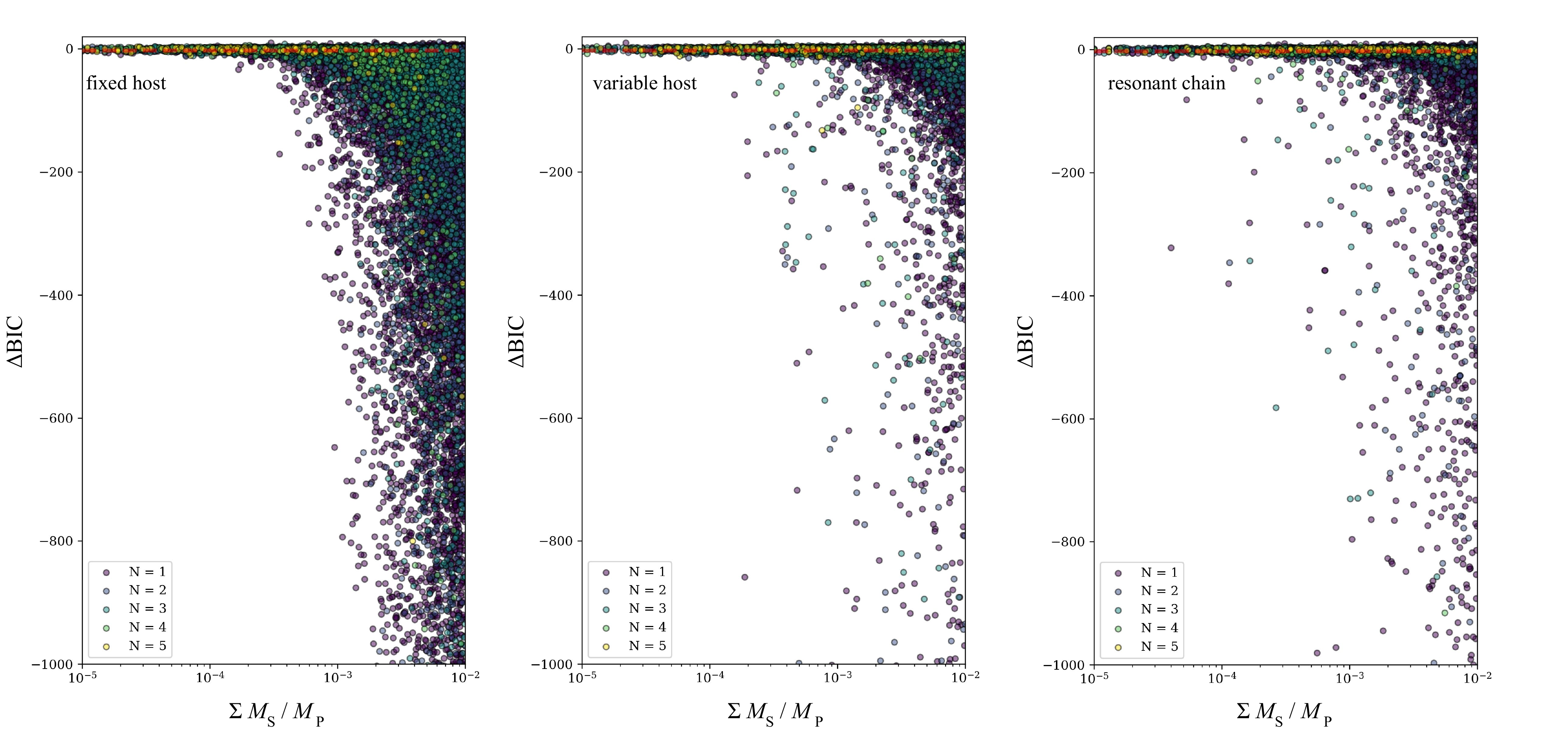}
    \caption{$\Delta$BIC as a function of total moon mass for the fixed host (left), variable host (center), and resonant chain (right) simulations, color-coded by the number of moons in the system. Our threshold for significance (-2) is indicated as a red dashed line.}
    \label{fig:deltabic_vs_moon_system_mass}
\end{figure*}

We turn now to examining the detectability of these signals. Figure \ref{fig:deltaBIC_histograms} indicates that the peak of each distribution falls right around $\Delta \, \mathrm{BIC} = 0$, but there are significant tails for these distributions in the fixed host systems, which can be responsible for pulling the median values well off the peak, especially for small $N$ systems. This tail is absent for the variable host and resonant chain systems, however. For the sample shown here we have implemented an additional, observational bias such that simulated planets are represented in proportion to their geometric transit probability. We have assumed any transiting planet is equally likely to be discovered, however; completeness is not a function of orbital period here.

We can see that as $N$ increases, there will be fewer and fewer such systems displaying strong evidence for TTVs based on the $\Delta$BIC metric. The systems also never display a more positive $\Delta$BIC than about 10, across all architectures, suggesting that this is what would be needed to strongly rule out the presence of TTVs based on our methodology.

Figure \ref{fig:deltabic_vs_moon_system_mass} offers a glimpse into what is happening here. As previously discussed, systems with fewer moons are in general more stable at higher mass ratios (taking their simulation survival rates as a proxy for stability). This in turn makes these systems on average more detectable, since a higher mass ratio corresponds to larger amplitude TTVs.


\subsection{Exomoon corridor}
\begin{figure*}
    \centering
    \includegraphics[width=17.7cm]{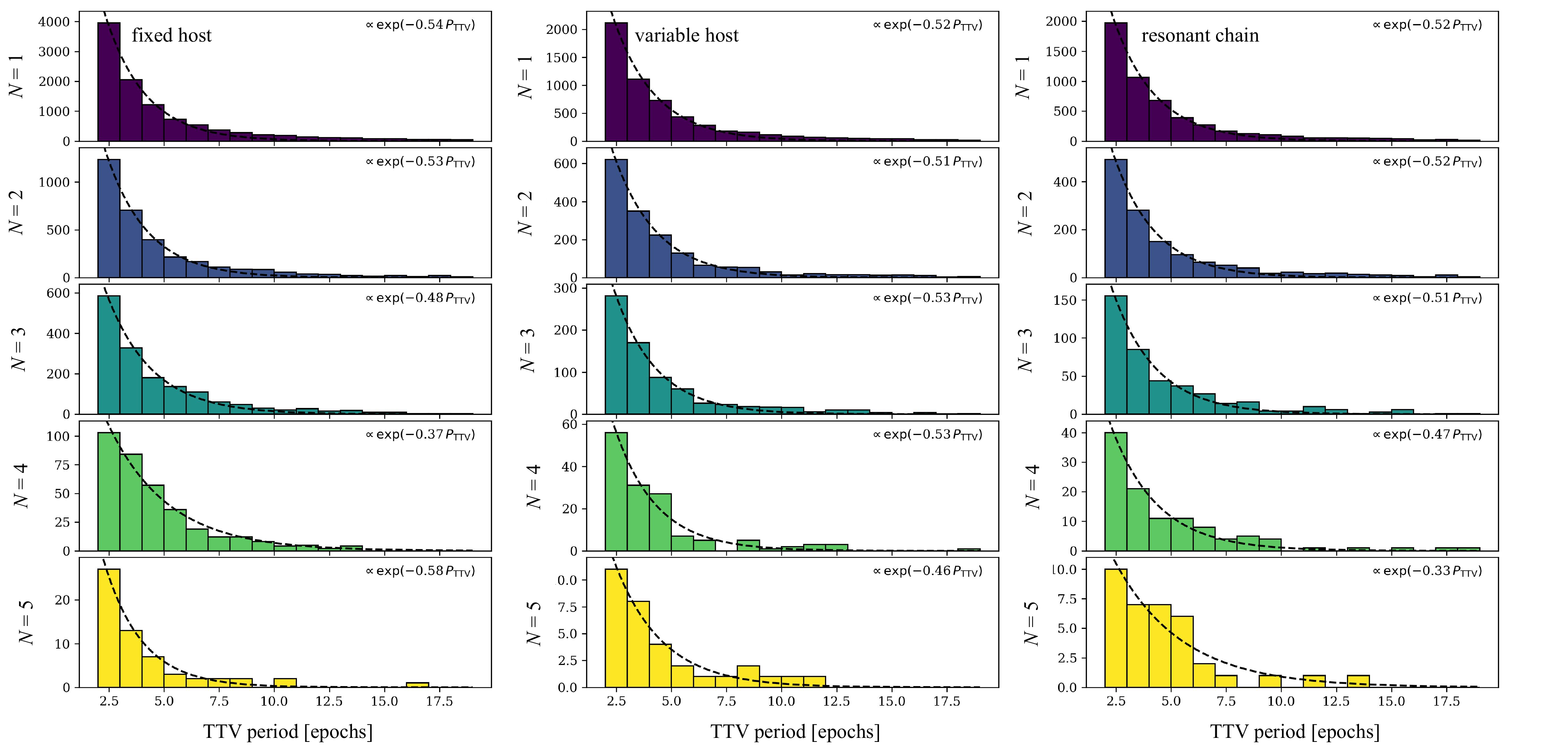}
    \caption{Distribution of inferred TTV periods for $1 < N < 5$ moon systems for the fixed host (left), variable host (center), and resonant chain (right) simulations. Results were similar when a geometric observational bias is applied, but suffer from smaller numbers in each group. These results demonstrate that the characteristic exomoon corridor distribution holds for $N > 1$.}
    \label{fig:distribution_of_TTV_aliases_by_Nmoon}
\end{figure*}

At the crux of this investigation is the question of whether the exomoon corridor finding -- the pile-up of short $P_{\mathrm{TTV}}$ values -- holds for $N \geq 1$ moons. Figure \ref{fig:distribution_of_TTV_aliases_by_Nmoon} demonstrates the distribution of measured TTV periods for the various moon architectures simulated in this work. 

We see that the exomoon corridor result does indeed generalize for $N \geq 1$ moons, and evidently this characteristic distribution of TTVs is insensitive to the number of moons in the system, showing remarkable consistency. Each system represented here has a $\Delta \mathrm{BIC} \leq -2$ in favor of TTVs over linear ephemeris, and survived our cross-validation test to ensure the period solutions are reliable. The short-period pile-up also occurs without the cross-validation cut, but it is important to screen out those systems with inconsistent period solutions to ensure that this is not merely a side-effect of our methodology. We also screen out systems for which the TTV period solution is precisely equal to 2, as these solutions are generally spurious, placing every epoch at either the beginning or end of the phase-folded solution and leaving a completely unconstrained curve in between. Applying the geometric observational bias yields virtually the same distribution as is shown here but suffers from a reduction in the number of systems available and is thus less clean.

\section{Analysis}
\label{sec:analysis}

\subsection{Comparing the three simulation samples}
In this work we have compared three different simulation runs: one with a restricted host system architecture (consisting of a Solar-analog star, Jupiter-analog planet, and a fixed photometric uncertainty), a second employing a realistic distribution of physical and observational attributes, and the last initializing all moon systems as resonant chains. Results from these runs are recorded in Table \ref{tab:summary}.

There is a great deal of consistency between the three sets of simulations. During the generation process we see quite similar numbers in terms of how many systems of each architecture survived random initialization. In particular, we note that all three $N=2$ systems had a survival rate roughly half that of $N > 2$ systems, suggesting that these systems are particularly prone to instability. We stress again that a careful interpretation is needed here; randomized $N=2$ systems are more unstable than the others, so the orbits of $N=2$ systems found in nature are unlikely to be ``random''.

Likewise the total satellite mass ratios, falling off with increasing $N$, follow the same trend across all three simulation runs. And we see that both the derived median $P_{\mathrm{TTV}}$ and $A_{\mathrm{TTV}}$ values are broadly similar across all three runs, as well as consistent across all choices for $N$. This is a very important result, because it means the exomoon corridor result holds regardless of the assumptions we build into our model. Whether the moons sit in resonant chains or not, the measured TTV periods will pile-up at the short end.

However, some important differences are evident. The total number of systems that show evidence for a TTV as measured by the $\Delta \mathrm{BIC}$ falls off precipitously from the fixed host to the variable host and resonant chain runs. This suggests than an occurrence rate for exomoons based on a dynamical test such as this, if such a rate can be derived, may be impacted by physical and observational system attributes. The situation is confusing, however, as the $\Delta$AIC metric does not display this same drop-off. This could be because our adopted thresholds for $\Delta$BIC and $\Delta$AIC are not strictly compatible.

Interestingly, the fixed host and variable host runs, where moon period ratios could take any value, show clear deficits at the 2:1 and 3:1 period ratios (Figure \ref{fig:period_ratios}), in keeping with expectations that resonances may contribute to instability. By contrast, the resonant chain simulations have peaks at these period ratios, but that is merely an artefact of the methodology for generating the moons for this run.

It is encouraging to see that the majority of systems for which $\Delta \mathrm{BIC} \leq -2$ also cleared the cross-validation test. This suggests that we are not imposing a significant additional bias in terms of the systems we examine; by and large a system with a detectable TTV will also maintain the best period solution even as data are stripped away.

\subsection{Stability, Detectability, and Limitations}
\subsubsection{implications for detectable systems}
As we have discussed, the simulation results (Figure \ref{fig:fraction_stable}) across all three simulations indicate a trend towards lower satellite system masses as we increase the number of moons. Multi-moon systems with higher masses can of course exist in nature -- assuming they can be formed in the first place -- but they will be comparatively rare, or they will have to be more finely tuned in order to survive.

Perhaps noteworthy is the fact that at $N=4$, system architectures become majority unstable when total satellite system mass ratios go much above $10^{-4}$ -- coincidental or not, this is in keeping with the result from \citealt{canup:2006}, who found a mass ratio limit of $\sim 10^{-4}$ based on simulations of moon accretion. At $N=5$, less than 2\% of the generated systems were predicted to be long-term stable. More recent work on circumplanetary disk (CPD) evolution and moon production \citep[e.g.][]{inderbitzi:2020, cilibrasi:2020} have similarly predicted generally lower mass satellite systems, which will in many cases be below our current dynamical detection limits.

We caution the reader against over-interpreting the present results; in particular, these stability fractions should not be read off as absolute occurrence rate probabilities or priors. Their initial configurations (number, distribution, and masses) are not dictated by CPD models; depending on the density profile of the disks, there may be more or fewer long-term stable moons produced in nature than what we produce here by virtue of a different initial distribution of moons. At the same time, it is worth keeping in mind that, due to the present dearth of exomoon system observations, CPD modeling has primarily (understandably) focused on explaining the regular moons of our Solar System \citep[e.g.][]{canup:2006, szulagyi:2018, Fujii:2020}, and therefore might conceivably reflect an observational bias in that regard.

Of course, moons are not only formed in CPDs; they may also be formed through impact scenarios \citep[e.g.][]{canup:2001, Canup:2005, Ida:2020} or capture scenarios \citep[e.g.][]{agnor:2006, nesvorny:2007}, for example. Generally, these formation events are typically expected to produce a single massive moon, with perhaps some much smaller moons as well, as in the case of Neptune \citep{banfield:1992} and Pluto \citep{canup:2011}. Still, it is at least conceivable that heretofore un(der)explored formation pathways give rise to systems unlike any seen in the Solar System \citep{TK18, Hamers:2018, Hansen:2019}. Thus, it is at least conceivable that we might find systems with more than one massive moon that did not form in a CPD, and we wish to simulate such systems. If they are stable, they could potentially be found in nature. So long as these systems are not vastly over-represented in the sample, and their inclusion does not wildly skew our results, we can feel confident in having them included.

Though our simulations are not produced from CPD profiles, they are broadly consistent with the output of CPD models in the sense that both predict we will not find many high mass, multi-moon systems. As such, we do not have to rely entirely on the predictions of the CPDs to infer the relative probabilities of high-mass multi-moon systems: even if the CPDs could produce them, the present analysis of post-CPD behavior suggests they will not be long-lived. To avoid this, the CPDs would have to be capable of fine-tuning the orbits.

If we accept that satellite systems with more moons will generally have lower masses, as the present results suggest, this points to a possibly important observational bias in detecting exomoons. Namely, we will be more likely to see systems with fewer moons, by virtue of the fact that they permit substantially higher mass ratios, and these higher mass ratios generally correspond to larger amplitude TTVs.

If CPDs are incapable of producing mass ratios in excess of a few times $10^{-4}$, we will then be biased towards identifying systems for which this was not the formation pathway; capture or giant impacts will be the more likely explanation for these massive moons, and if the Solar System is any guide, these moons will either be alone (as with the case of our Moon), or they will dwarf their companion moons in the system (as with Triton and Charon). It is of course not currently known how likely these scenarios are in exoplanetary systems; if these pathways are less common, exomoon detections may be quite rare in the near term. In any case, great care will need to be taken in interpreting the relative occurrence rates of these types of satellite systems as a window into planetary system formation.

Where will we find these moons? On the one hand, shorter period planets will display more transits, and this is clearly advantageous as we search for dynamical signatures: more transits mean more data points with which to work, and a clearer signal can emerge. On the other hand, longer period planets have their own advantages. They will generally have longer transit durations, which boosts the timing precision (though this is a weak effect, as the transit SNR only scales with $\sqrt{n}$ in-transit data points). Moreover, as we just pointed out, planets on wider orbits may have corresponding larger Hill spheres, allowing for moons on wider orbits, and greater durability against moon-moon scattering events. The occurrence rate of these wide-separated moons, of course, can be modeled but is not yet known empirically.

\subsubsection{stability proxies}
It is worth remembering that the stability fractions presented in Figure \ref{fig:fraction_stable} and Table \ref{tab:summary} are based on short-integration stability predictions, and for $N>3$ systems this prediction is based on \texttt{SPOCK}, which was not originally designed for moon systems. In particular, the \texttt{SPOCK} training set did not include mass ratios up to $10^{-2}$ as included here. The code works by utilizing a number of metrics shown to be strongly predictive, and based on its excellent performance for non-resonant systems and for systems with $N>3$ planets (i.e. both also outside the training set), \texttt{SPOCK} is expected to be reliable in this higher mass regime (D. Tamayo, private communication). But we stress that this has not been fully demonstrated. We proceed under the assumption that \texttt{SPOCK} is performing well enough in this regime for our purposes of producing plausibly observed systems.

Along these lines, we remind the reader that the MEGNO number was utilized for predicting stability for $N=2$ systems. Our range of acceptable values for MEGNO was conditioned on \texttt{SPOCK} predictions for $N \geq 3$ systems, so strictly speaking we are comparing apples and oranges here. A more stringent limit on the acceptable MEGNO range would certainly reduce the number of $N=2$ systems considered stable, and this should be taken into account when considering the number of surviving $N=2$ systems shown in Figure \ref{fig:fraction_stable}. 

\subsubsection{model limitations}
We caution the reader that we have not explored the full range of possible system architectures, including various orientations with respect to the host star / planet's orbit, mutual inclination of the moon orbits, or non-spheroidal geometry of the planet's gravitational influence. These would undoubtedly affect the resulting stability maps, likely reducing the total number of stable systems across the board. Notably, exploring the \textit{full} range of architecture possibilities would itself be a departure from CPD model predictions and expectations based on the Solar System. We could explore moons with large mutual inclinations, for example, but this architecture is \textit{a priori} less likely to be found in nature than systems that are essentially co-planar with each other (as we seen with the major moons of Jupiter, Saturn, and Uranus). 

Utilizing a realistic distribution of planet and star masses (as was used for both the variable host and resonant chain simulations) has both observational and dynamical consequences. Dynamically, the primary influence these variables will have is on the size of the planet's Hill sphere, which is proportional to $\left( \frac{M_P}{ 3M_*}\right) ^{1/3}$. Star and planet sizes were drawn independently, but because the latter is based on a real detected transit, the mass of the planet is conditioned on the mass of the star. That is, planets simulated for larger stars will be on average larger than planets produced for smaller stars, mirroring both physical and observational correlations in the real \kepler\ sample. We did not attempt to account for possible differences in planet distributions as a function of spectral type, but we note that we drew uniformly in orbital period, not in semimajor axis, so there will be a non-uniform distribution in $a$ due to the mass distribution of the host stars. 

The size of the planet's Hill sphere will of course not only dictate the maximum semimajor axis of a stable moon; it will also determine how tightly packed moons may be. For a given moon architecture and mass ratio, a larger Hill sphere will naturally afford more ``breathing room'' for the satellite retinue. A wider berth for each moon will translate to weaker dynamical interactions between the moons and generally more stable systems. It is worth keeping in mind however that if this breathing room goes unused by planets on wide-orbits in nature, such that moons of longer-period planets tend to reside at smaller fractions of the Hill radius than moons around shorter-period planets, this breathing room effect will be mitigated.

Even so, our simulations do not show any relationship between the planet's period (as a proxy of the planet's Hill radius) and the degree of stability. But this is due to the way the moon architectures were built, to optimize for survivability, dividing up the stability region more or less uniformly between the moons, and each architecture getting the full range of possible mass ratios. With such an approach to constructing the systems, the stability of these systems will be determined by the masses of the moons themselves (lower mass moons clearly produce smaller gravitational perturbations on their neighbors), and that is precisely what we see in our simulations.

\subsection{On the cross-validation test}
We should consider the question of whether the cross-validation test utilized here to ensure robustness of the $P_{\mathrm{TTV}}$ measurements might be skewing our results. It could be argued that by insisting on a single, coherent, detectable signal even as we strip away data points, we are effectively asking this system to resemble a single moon system, and therefore we are virtually guaranteed to recover the single moon exomoon corridor result. 

The short answer is yes, in a way that is precisely what we are doing; a single, coherent signal is a prerequisite for calling the measured TTV signal ``real''. If this signal vanishes, for whatever reason, it is rejected. But because our aim is to determine whether the exomoon corridor finding holds for $N > 1$ moons -- which is by no means guaranteed -- we have to test whether systems with $N > 1$ systems display the same behavior when applying the exomoon corridor methodology, which implicitly assumes a single-moon model manifesting a single sinusoid. At the same time, we also require some methodology for ensuring that the $P_{\mathrm{TTV}}$ we measure for a real system is not due to bad data points which skew towards high frequency solutions. And we want to make sure that the period solution is not a function of the observing cadence; it should be robust against changes in the signal sampling.

Recall, too, that most systems we generated, regardless of the architecture, yielded a cross-validated solution, so it is not the case that we are screening out all but the most advantageous systems. In light of this, we might think of the cross-validation test as a kind of false alarm probability (FAP) test, ensuring that the signal we are measuring is robust against outliers. The cross-validation has an advantage over the FAP test, however, in the sense that it is conditioned on each individual light curve's $P_{\mathrm{TTV}}$ solution, rather than being a tuned parameter that is conditioned on the entire dataset. Since the observing cadence, and deviations from uniform sampling, will be unique to each planet, a metric tailored to each planet may be preferrable.

For which systems is this approach appropriate, and which are the problem cases? Generally speaking, planets with shorter orbital periods (and correspondingly more transit observations), will be more robust against the removal of data points. With the method adopted here, no fewer than $5\%$ of the observations will be left out for any cross-validation test, but for longer period planets this can be a much higher percentage. In the limit of having a planet with only four available transits (the minimum required to leave out one and still measure a TTV with the remaining observations), leaving out a single data point removes a full quarter of the information. That can have a significant impact on the inferred periodicity of the TTV (though this is hardly enough to put meaningful constraints on a model fit anyway). These systems have a good chance of being rejected during the cross-validation process.

Longer moon periods do not translate to longer inferred moon periods from the periodogram. We refer the reader again to Figure \ref{fig:PTTV_aliases}, which shows the aliased TTV period as a function of moon and planet orbital periods (implementing equation 10 from K21). For a given satellite period, the inferred $P_{\mathrm{TTV}}$ will rise and fall repeatedly based on the orbital period of the planet (that is, the sampling cadence). There are therefore infinitely many possibilities for any given moon period or planet period, though there is a well-defined solution when both the planet and moon periods are known. The introduction of timing uncertainties will complicate this clean picture, as will a non-uniform sampling, and of course, the introduction of additional moons.

The Lomb-Scargle periodogram was designed and is used for the expressed purpose of producing a power spectrum for a signal with non-uniform sampling. Nevertheless, non-uniform sampling is a major contributor to spurious (non-astrophysical) power in the spectrum. The simulated systems produced in this work all started with uniform sampling, but with the cross-validation test we have in fact simulated non-uniform sampling through random removal of data points, and thus produced such confounding signals in the periodogram which do in fact arise with real data. This is therefore also an important test for robustness of the moon corridor result as applied to real data. Because the large majority of systems observed by \textit{Kepler} have at least one transit missing, we have to contend with the spurious power in the periodogram and want to know that the exomoon corridor holds up even in these cases.

\subsection{Field comparison}
Having satisfied ourselves that the exomoon corridor result holds for $N \geq 1$ moons, our last task is to examine the real \kepler\ sample and the HL2017 subset examined by K21. The value of the exomoon corridor result rests in part on the expectation and observation that it provides some discriminating power for identifying promising signals, so it is worth testing this further. From a theoretical standpoint the discriminating power should be a given; once the signal is no longer under-sampled, with a sufficient time baseline it should be possible to measure the true super-period. Planet-planet perturbations, which certainly do produce much lower frequency oscillations \citep{Lithwick:2012}, will generally not suffer the undersampling issue present for moons. The super-period will be longer than the orbital period of the planet, unlike the oscillation period of the planet due to the moon.

This is not to say that only moons may produce an undersampled signal. For example, in the case where an inner planet perturbs the star to such an extent that the light source for the outer transiter is no longer effectively stationary, this too would produce a oscillation signal of higher frequency than the observing cadence, as the inner planet's period must be shorter than the outer transiter.

Naturally, we also want to investigate whether there is any evidence for moons in the data on hand based on this approach. We therefore proceed with a comparison to the real data.

\begin{figure*}
\includegraphics[width=\textwidth]{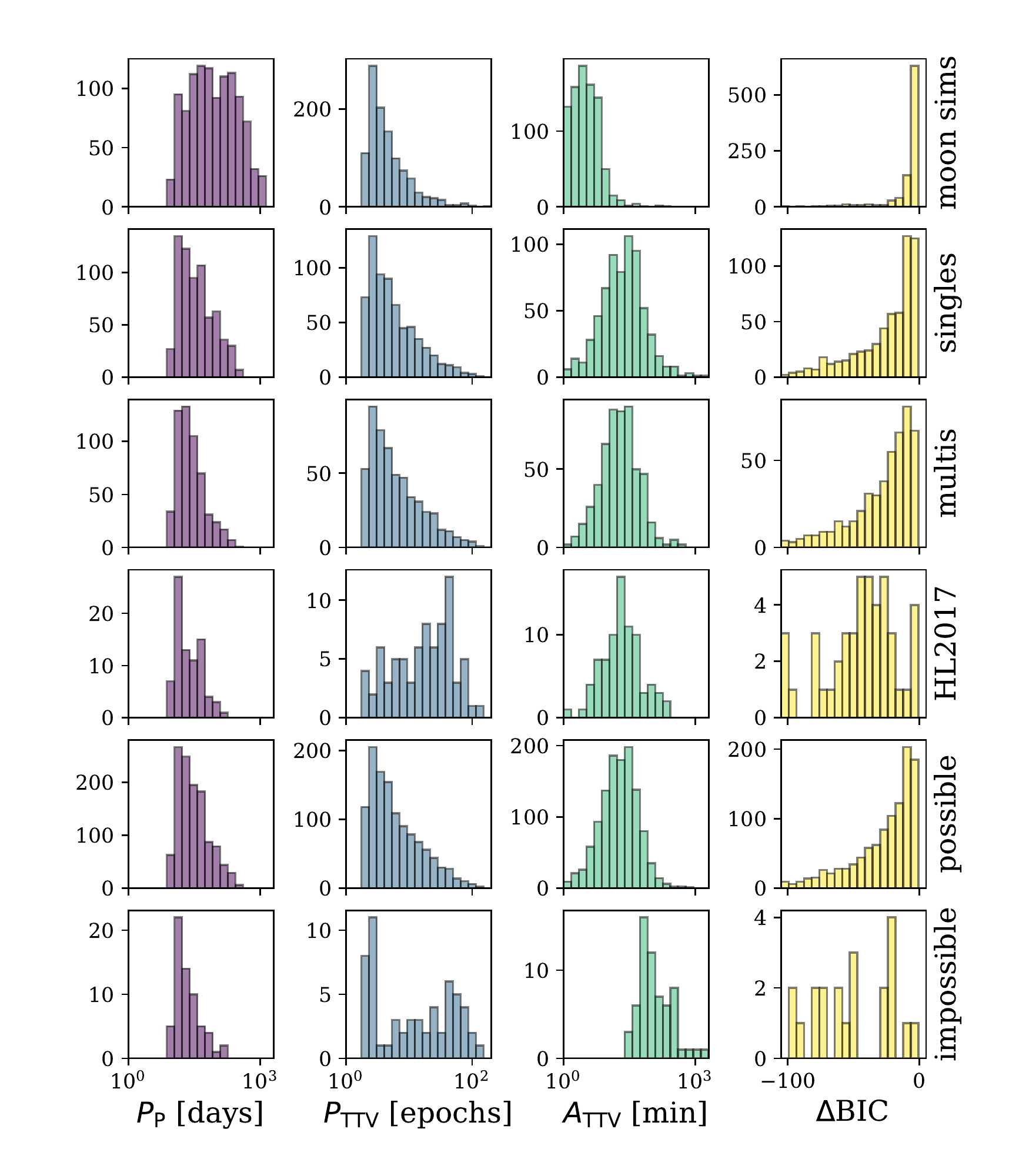}
\caption{Distribution of planet periods, TTV periods, amplitudes, and $\Delta$ BIC values for the simulations in this work (resonant chain run), single-planet systems with TTVs from \citealt{holczer:2016}, multi-planet systems with TTVs from the same work, multi-planet systems taken from HL2017, ``possible'' moons and ``impossible'' moons. All real planets with periods shorter than 10 days have been excluded.}
\label{fig:6x4}
\end{figure*}

For the \textit{Kepler} TTVs we utilized the catalog from \citealt{holczer:2016}. False positives (based on the NASA Exoplanet Archive designation) were screened out, and each planet was subjected to the same methodology as the simulations, namely, screening outliers with \texttt{DBSCAN}, running a periodogram to infer the $P_{\mathrm{TTV}}$, testing that solution's significance, and performing the cross-validation test. We further screened out any systems for which the best $P_{\mathrm{TTV}}$ solution was precisely equal to 2, as this is virtually always a bad fit, with all data points at the nodes and a completely unconstrained amplitude.

We can further examine the HL2017 subset, consisting of 90 planet pairs, in an attempt to replicate the results of K21 with our methodology, and to determine the degree to which this set represents the broader sample of \kepler\ planets. Establishing whether the HL2017 planets are a biased sampling from \kepler\ is a considerably more difficult question to answer, though we may attempt to tease out some answers.

Figure \ref{fig:6x4} shows histograms of relevant quantities for the simulations, and various cuts of the \citealt{holczer:2016} sample: 1) ``single'' planets, for which no other planets have been discovered in the same system; 2) known multi-planet systems; 3) the HL2017 subset, consisting of 90 planet pairs before quality cuts; 4) systems for which a moon can plausibly explain the observed TTV amplitudes (based on the \citealt{impossible_moons} formulation); and 5) systems for which a moon is insufficient to explain the observed TTV amplitudes. We discuss each of these subsets in turn in relation to the simulated systems produced in this work.

\subsubsection{``single'' planet systems}
It is important to keep in mind that the so-called ``single'' planets may or may not actually be alone; we simply do not \textit{know} of other planets in the system. There could be non-transiting planets, interior or exterior to the planet in question, that are capable of producing these TTVs. That is to say, the exomoon hypothesis need not necessarily be the explanation for the TTVs. But certainly \textit{something} is producing the TTVs, and exomoons are one plausible solution. 

For these systems, we see the characteristic pile-up at short period TTVs, as predicted by the exomoon corridor. This could suggest that many of these systems are potential moon hosts, but we must temper this reading based on the results of the other cuts discussed below. We also point out that the TTV amplitudes are on average larger than those of the simulations, but there is significant overlap between the two distributions. The TTV amplitudes are important to keep in mind, because they are indicative of both moon mass and semimajor axis, the latter of which will be limited by the planet's proximity to the host star. Smaller TTV amplitudes will be generally speaking more moon-like.

The $\Delta$BIC distribution also has a much longer tail than the simulations. Both of these results (larger amplitudes and more negative $\Delta$BICs) apply also to all but the ``impossible'' moons cut described below, where the $\Delta$BIC tail is also present but there is very little overlap between the TTV amplitude distributions. 

\subsubsection{multi-planet systems}
For multi-planet systems, a TTV may be plausibly explained by the presence of one or more other planets in the system. A thorough analysis of the probability of this explanation for each individual system, based on the capacity for other known planets in the system to induce the observed TTVs, is beyond the scope of this work. But the presence of neighboring planets certainly does not guarantee strong TTVs. And moons are clearly still possible in such systems; a planet may feel both planet-planet perturbations as well as planet-moon perturbations. 

Figure \ref{fig:6x4} shows that multi-planet systems in the \citealt{holczer:2016} catalog also show a pile-up at short TTV periods. This is a bit vexing, because we expect at least some of these systems to have TTVs induced primarily by neighboring planets rather than exomoons, which we predict will have generally higher $P_{\mathrm{TTV}}$ values. Still, because we have weak constraints on the occurrence rates of exomoons, we cannot observe this distribution and conclude that moons are or are not present in large numbers, nor can we conclude that the exomoon corridor result has little predictive power. It may very well be that planets with both planet- and moon-induced TTVs will have the inferred signal dominated by the latter, but this remains to be shown.

\subsubsection{HL2017}
The fourth row of Figure \ref{fig:6x4} shows the planets from the HL2017 catalog, for which the measured TTVs may be reliably attributed, at least in part, to planet-planet interactions. We note that the distribution plotted here is somewhat different from that shown in K21, and this is due to differing methodologies: the HL2017 subset presented here simply utilizes the \citealt{holczer:2016} TTVs for this selection of planets, processed with the same pipeline as all the others. K21 on the other hand computed the TTVs analytically from system attributes published in HL2017. 

Even so, we still see that the peak of this $P_{\mathrm{TTV}}$ distribution is well to the right of the exomoon corridor pile-up, consistent with the K21 analysis, suggesting the exomoon corridor may indeed have some discriminating power. On the other hand, the distribution presented here also shows a minority of samples with shorter period TTVs, within the exomoon corridor but not showing the characteristic shape. Of course, we should be cautious about drawing strong conclusions, as we are dealing with a comparatively small set of planets in this sample.

What do we make of this distribution? By this methodology anyway, the distributions of moon-induced and planet-induced TTVs are not so cleanly separated as they are in K21. It bears repeating that just because TTVs are due to a perturbing planet (as they are in the HL2017 sample) does \textit{not} imply the absence of one or more moons. Moreover, considering the fact that the $P_{\mathrm{TTV}}$ distribution in K21 is derived analytically, there could be additional signals present in the data that are simply not captured in the analytical framework. That is, the K21 distribution might be thought of as a kind of cleaned $P_{\mathrm{TTV}}$ distribution, for which the planets are unambiguously the cause, without the confounding features of real data. As such, the K21 may be idealized, but shows more convincingly how the exomoon corridor can work in theory. It is too early to say whether that is the case, but at this point we should not be overly concerned about one methodology or the other \textit{vis-a-vis} any discrepancies between the resulting distributions.

\subsubsection{``possible'' moon systems}
We turn now to look at the ``possible'' and ``impossible'' moon systems, based on the \citealt{impossible_moons} framework. For the ``possible'' systems, where a moon can potentially explain the observed TTV amplitudes, the now-familiar distribution at short $P_{\mathrm{TTV}}$ is unmistakable.

We further point out in Figure \ref{fig:frac_Rhill} that among the systems appearing in the lower, ``possible moons'' portion of the plot (in green, below the dashed line), most of these systems are designated as singles. As such, much of the analysis above relating to single-planet systems applies here as well. These results are all consistent with the presence of moons, but could potentially be explained by the presence of an unseen planet. As before, these two possibilities are not mutually exclusive.

\subsubsection{``impossible'' moon systems}

The last row in Figure \ref{fig:6x4} depicts the so-called ``impossible'' moon systems, where a physically plausible moon cannot sufficiently explain the observed TTVs. For these systems we expect a distribution more like that of the HL2017 sample, and that is indeed what we see: a peak at significantly longer periods, but also, a good fraction of systems with shorter periods, as well. Like the HL2017 subset, this sample suffers from having small numbers. A large fraction of the ``impossible'' moon planets in the \citealt{holczer:2016} catalog have periods shorter than 10 days, but we have excluded those to be consistent with the simulations.\footnote{We note that recent work from \citealt{Dobos:2021} finds a majority planets with $P \lesssim 10$ days have no stable moon orbits, so this cut seems to be reasonable.}

As with the multi-planet systems discussed above, we must remember that a moon is certainly not ruled out by this TTV amplitude test; it is simply \textit{insufficient} to explain the observed TTV amplitudes. A perturbing planet is almost certainly responsible for these TTVs, at least in part. 

\begin{figure}
    \centering
    \includegraphics[width=8.5cm]{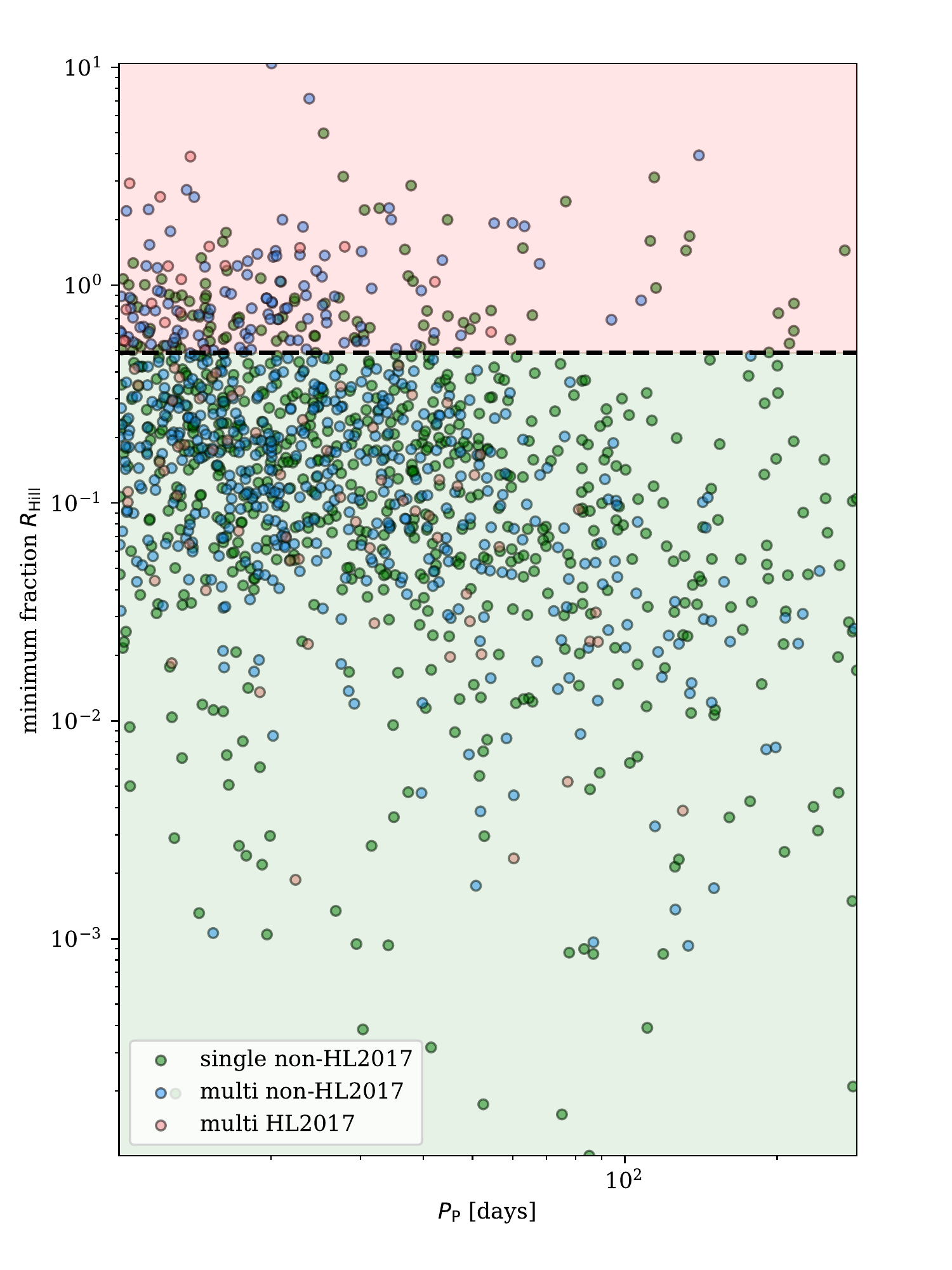}
    \caption{Minimum fraction of the Hill radius required of a moon in order to explain the observed TTV amplitude as a function of the planet's period. Objects above the dashed line (red shaded region) are ``impossible'' moons, that is, a moon cannot sufficiently account for the observed signal. Objects below the dashed line could potentially be explained by a moon. Objects are color-coded as single planet systems (green circles), multiplanet systems in the HL2017 catalog (red circles) and not in the catalog (blue circles).}
    \label{fig:frac_Rhill}
\end{figure}

Notably, Figure \ref{fig:frac_Rhill} also demonstrates that there is a significant number of singles in the impossible moon region (upper portion of the plot), showing that indeed several of these singles are not really alone. As a point of comparison, according to the NASA Exoplanet Archive there are currently only 20 planets in all the \kepler\ systems that have been discovered by radial velocity methods (that is, they were not seen transiting), which suggests our completeness for non-transiting companions remains quite poor. It is a good bet that most of the ``singles'' do in fact have companions.

In any case, it is perplexing that, given that the amplitude of the TTVs are such that a planet-planet interaction must be at least partly responsible, a short-period solution (more ``moon-like'') is still preferred in a sizeable fraction of cases. Naively we should expect that the largest signal in a multi-signal time series should have the most substantial power, so the best period solution ought to correspond to the planet-planet interaction with an (on average) longer $P_{\mathrm{TTV}}$. After all, these systems are in the ``impossible'' moon set precisely because their amplitudes are too large to be moon-induced, so we would expect the $P_{\mathrm{TTV}}$ solution to accord with that.

Of course, the large TTV amplitudes that place these systems in the ``impossible'' moon category could be due to the superposition of a moon signal and a planet signal, where each individually might come in under the threshold. Or they might even be due to the superposition of multiple moon signals. It is also conceivable that these large amplitudes are merely due to very large timing uncertainties, which will generally be associated with lower SNR planets -- often short period planets, which have shorter transit durations and are generally detectable with shallower transits due the availability of more transits for stacking. And as previously mentioned, perturbations of the light source by a (possibly non-transiting) interior planet could also be responsible for large amplitude, under-sampled TTV signals. The relative likelihood of this scenario, given the low occurrence rate of Hot Jupiters \citep[e.g.][]{wright:2012, howard:2012} would have to be considered, however. 

\subsubsection{utility and limitations of the exomoon corridor framework}

In considering the results presented in Figure \ref{fig:6x4}, we may be tempted to conclude that the very moon signal we seek is making itself plain to us in the data. After all, we do not have exceptionally firm numbers on the exomoon occurrence rate, so it remains plausible that there is an abundance of yet-undetected moons hiding in the data, showing up in the exomoon corridor. Indeed, as a dynamical indicator, it can potentially reveal moons that may be missed by looking for transits, as in \citealt{HEKVI}. A thorough accounting of TTVs that may be explained by known planets, or not \citep[cf.][]{Kane:2019}, could help identify systems in which the moon hypothesis becomes a more likely scenario.

On the other hand, we must acknowledge the possibility that, despite our best efforts, we have still not sufficiently screened out spurious (outlier) TTV data points which, as we've discussed, can have the effect of driving the period measurements downward, mimicking the characteristic exomoon corridor distribution. This is a very plausible explanation for the pile-up, but rather difficult to test, as we must be careful not to massage the data to the point of achieving the desired result. We have attempted to make reasonable choices for outlier rejection and verified the methodology by eye on a sample of systems. 

The exomoon corridor pile-up seen for the singles, multis, and possible moon systems might at first glance suggest that perhaps the methodology adopted in this work is simply always reproducing the exomoon corridor distribution, potentially undermining our result showing the exomoon corridor generalizes for $N \geq 1$ moons. Fortunately, the recovered distribution of $P_{\mathrm{TTV}}$ values for the HL2017 subset roughly reproduces the distribution computed analatically in K21, and the ``impossible'' moon systems also show a markedly different morphology. This is encouraging; if our code were simply always biased towards the short periods, it would presumably show up for these subsets also. 

Even so, the excess distribution of TTV periods at the low end of the HL2017 sample compared to the distribution shown in K21 might suggest that the quality of the TTVs from \citealt{holczer:2016} may be partly responsible for these results, and that a more aggressive removal of bad data or systems could yield a cleaner result. Or, as we've said, it may be that the K21 distribution reflects only planet-planet perturbations and misses the additional power in the TTVs due to its analytical approach.

All these possibilities deserve further investigation in future work. Suffice it to say, the distribution of $P_{\mathrm{TTV}}$ values for the multi-planet systems, and for some fraction of the HL2017 and ``impossible'' moon systems (those at short periods), raise some questions about the utility of the exomoon corridor result in identifying promising moon systems using real data. All the same, these investigations do not negate the fundamental findings of K21 or this work.

The question remains, is the HL2017 sample biased in such a way that makes longer period TTV solutions more likely than the rest of planet-planet TTV systems? That is difficult to answer. Table \ref{tab:kstest} shows $p-$values from a Kolmogorov-Smirnov (K-S) test comparing the HL2017 sample to all the planets in the TTV catalog and for the multi-planet system subset. We find that while the HL2017 subset is fairly representative in terms of the TTV amplitudes and marginally representative of the KOIs in the \citealt{holczer:2016} sample in terms of the planet periods, they are quite distinct from the multi-planet systems in terms of $P_{\mathrm{TTV}}$ and planet radius. This by itself does not demonstrate a bias, but it suggests we must be careful about considering the HL2017 sample a proxy for the broader sample of \kepler\ multis. 

These questions will have to be resolved before we can interpret the exomoon corridor result pile-up in the \kepler\ sample as showing evidence for exomoons. The findings are certainly \textit{consistent} with abundant exomoons in the data, but a deeper investigation of other possible explanations for this pile-up will be necessary.

\begin{table}
  \centering
  \begin{tabular}{|c|c|c|}
    \hline
    \textit{Quantity} & \textit{HL2017 vs all} & \textit{HL2017 vs multis} \\
    \hline
    $P_{\mathrm{P}}$ & 0.01 & 0.07  \\
    \hline
    $P_{\mathrm{TTV}}$ & $2.9 \times 10^{-14}$ & $4.1 \times 10^{-12}$ \\
    \hline
    $A_{\mathrm{TTV}}$ & 0.92 & 0.59  \\
    \hline
    $R_{\mathrm{P}}$ & $5.0 \times 10^{-4}$ & $5.3 \times 10^{-5}$ \\
    \hline
    
  \end{tabular}
  \caption{$p$-value results of K-S test comparison of HL2017 sample to the \kepler\ population. The value represents the probability that the two samples are drawn from the same distribution.}
  \label{tab:kstest}
\end{table}


\section{Conclusions}
\label{sec:conclusions}
In this work we have examined the generalization of the exomoon corridor result to systems with $N \geq 1$ moons. We carried out $N$-body simulations of 150,000 planet-satellite systems, computing long-term stability probabilities, and found that the characteristic distribution of measured TTV periods presented in K21 does in fact hold in the more general case, demonstrating the potential of the approach for identifying possible exomoon host systems with TTV signatures alone. These results hold for both resonant and non-resonant chain systems and should therefore apply to virtually any multi-moon system found in nature.

In the course of our simulations we found that systems with more moons are on average required to have significantly smaller total satellite mass ratios, or their architectures must be more finely tuned, in order to survive. This fine tuning will have to be achieved during the the circumplanetary disk phase. For systems with $N \geq 4$ moons, the majority of systems were long-term unstable for mass ratios exceeding $\sim10^{-4}$ -- systems which rarely present detectable dynamical signatures -- suggesting that we will be biased against seeing multi-moon systems and thus also against finding systems formed in a circumplanetary disk; impact and capture scenarios will be more common formation pathways for exomoons discovered in the near term through dynamical means. This may have bearing on any statistical inferences made regarding planetary system dynamics based on the presence of moons. 

Finally, we compared our simulation results to the real \kepler\ sample using the \citealt{holczer:2016} catalog to examine the extent to which the exomoon corridor pile-up occurs in the field. We found the same exomoon corridor pile-up at short $P_{\mathrm{TTV}}$ for both ``single''-planet systems and multi-planet systems, as well as systems for which a moon can plausibly explain the TTV amplitude. These results are consistent with the presence of moons, but it is premature to interpret this distribution as \textit{evidence} of moons. More work is needed to determine whether the $P_{\mathrm{TTV}}$ pile-up can be sufficiently explained without an abundance of moons in the data. Meanwhile, the HL2017 sample, and the ``impossible'' moon sample, show markedly different distributions, suggesting that the exomoon corridor does have some discriminating power, though these samples both suffer somewhat from small sample sizes. Taken together, these results are intriguing but also suggest some lingering challenges in utilizing the exomoon corridor as a tool for moon hunting.

In our Solar System, multi-moon systems are the norm, not the exception, and we have reason to anticipate the same may be true for exoplanets, as well. Despite the remaining challenges of detecting and demonstrating their presence, they will remain an important class of systems that we should continue to investigate theoretically and observationally in the years to come.


\section*{Acknowledgements}
We thank the anonymous referee for their constructive feedback which improved this paper. We thank Daniel Tamayo for his suggestions relating to the implementation of \texttt{SPOCK}, and David Kipping with valuable discussions about the exomoon corridor result. This research has made use of the NASA Exoplanet Archive, which is operated by the California Institute of Technology, under contract with the National Aeronautics and Space Administration under the Exoplanet Exploration Program. This research has made use of NASA's Astrophysics Data System. This paper includes data collected by the \kepler\ mission and obtained from the MAST data archive at the Space Telescope Science Institute (STScI). Funding for the \kepler\ mission is provided by the NASA Science Mission Directorate. STScI is operated by the Association of Universities for Research in Astronomy, Inc., under NASA contract NAS 5–26555. We also gratefully acknowledge the developers of the following software packages which made this work possible: \texttt{Astropy} \citep{astropy1, astropy2}, \texttt{NumPy} \citep{numpy1, numpy2}, \texttt{Pandas} \citep{pandas}, \texttt{Matplotlib} \citep{matplotlib}, and \texttt{SciPy} \citep{scipy}.

\section*{Data Availability}
The code used to produce the simulations and perform the analysis are hosted at \href{https://github.com/alexteachey/Nmoon_TTVs}{https://github.com/alexteachey/Nmoon\_TTVs}. Simulations produced in this work amount to $\sim$15GB of data and are stored on a hard drive belonging to the author, requests for access may be made via e-mail. The \citealt{holczer:2016} TTV catalog utilized for the \kepler\ field comparison may be found at \href{ftp://wise-ftp.tau.ac.il/pub/tauttv/TTV/ver_112}{ftp://wise-ftp.tau.ac.il/pub/tauttv/TTV/ver\_112}.












\bsp	
\label{lastpage}
\end{document}